\documentclass[12pt]{iopart}

\newcommand{\be}{\begin{eqnarray}}
\newcommand{\ee}{\end{eqnarray}}
\newcommand{\bea}{\begin{eqnarray}}
\newcommand{\eea}{\end{eqnarray}}

\usepackage{graphicx}
\usepackage{bm}
\usepackage{color}
\usepackage{slashed}
\usepackage{cite}
\usepackage{graphicx}
\usepackage{multicol}
\usepackage{graphicx}
\usepackage{color}
\usepackage{slashed}
\usepackage{CJKutf8}

\begin{document}
\begin{CJK}{UTF8}{<font>}
\title{Shadow thermodynamics of AdS black hole in regular spacetime}

\author{Sen Guo$^{1}$, \ Guan-Ru Li$^{1}$, \ Guo-Ping Li$^{2*}$}

\address{
$^1$Guangxi Key Laboratory for Relativistic Astrophysics, School of Physical Science and Technology, Guangxi University, Nanning 530004, People's Republic of China\\
$^2$School of Physics and Astronomy, China West Normal University, Nanchong 637000, China}

\ead{sguophys@126.com; 2007301068@st.gxu.edu.cn; gpliphys@yeah.net}
\vspace{10pt}
\begin{indented}
\item[]May 2022
\end{indented}

\begin{abstract}
The dependence of the black hole (BH) shadow and thermodynamics may be structured in the regular spacetime. Taking the regular Bardeen-AdS BH as an example, the relationship between the shadow radius and the event horizon radius is derived. It is found that these two radii display a positive correlation, implying that the BH temperature can be rewritten as a function of shadow radius in the regular spacetime. By analyzing the phase transition curves under the shadow context, we found that the shadow radius can replace the event horizon radius to present the BH phase transition process, and the phase transition grade can also be revealed by the shadow radius, indicating that the shadow radius may serve as a probe for the phase structure in the regular spacetime. Utilizing the temperature-shadow radius function, the thermal profile of the Bardeen-AdS BH is established. We obtained that the temperature shows an N-type change trend in $P<P_{\rm c}$ situation. These results suggest that the phase transition process of the regular AdS BH can be completely presented in the thermal profile, and the relationship between the BH shadow and thermodynamics can also be established in the regular spacetime.
\end{abstract}

\noindent{\it Keywords}: Black hole shadow; Thermodynamics; Regular spacetime

\section{Introduction}
\label{intro}
\par
Over the last several decades, abundant astronomical evidence for black holes (BH) has accumulated from various sources. The observations of gravitational waves emitted from BH mergers by the Laser-Interferometer Gravitational Wave-Observatory (LIGO) is the first evidence of the existence of BHs from astronomical observations \cite{1}. The Event Horizon Telescope (EHT) reports an image of the supermassive BH in M87$^{*}$, convincing direct evidence of BH's existence in our universe \cite{2,3,4,5,6,7}. The major feature of this image is that the BH event horizon is surrounded by a dark area called BH shadow and a bright ring-shaped lump of radiation surrounding BH shadow. Based on the strong gravitational lensing, the formation mechanism of the BH shadow is that the specific photons around a BH collapse inward to produce the shadow, indicating the shadow can reflect the information of the jets and matter dynamics around the compact objects \cite{8}. The shadow can limit the mass, the spin, the charge, and other physical parameters of the BH, providing abundant sources of data information for finding new gravity theories \cite{9,10,11,12}.

\par
The most striking feature of a BH - namely, the event horizon - is only indirectly inferred. The BH shadow as an observable quantity, which may replace the event horizon when displaying some physical properties of BHs. The BH as a thermodynamic system is similar to the classical thermodynamic systems. For the BHs in AdS spacetime, the charged AdS BHs are almost equivalent to the van der Waals (vdW) liquid-gas system \cite{13,14,15,16,17,18}. The BH is not only regarded as a thermodynamic system but also as a strong gravitational one. Hence, it is necessary to establish the connection between its thermodynamics and dynamics. Wei $et~al.$ found that the unstable circular orbital motion can connect the BH phase transitions by establishing the relationship between the null geodesics and thermodynamic phase transition for the charged AdS BH \cite{19}. In Ref. \cite{20,21}, the authors deemed that the radius of a test particle's time-like or light-like circular orbital motion can be used as a characterized quantity of the BH phase transition information.

\par
Based on these researches, a fundamental connection between the BH thermodynamics and shadow is proposed. Zhang $et~al.$ found that the shadow radius can reflect the BH phase structure \cite{22}. By investigating the relations between the shadow and critical behavior of the charged AdS BH, Belhaj $et~al.$ obtained the critical exponent that perfectly matched the vdW system \cite{23}. Cai $et~al.$ studied the relationship between shadow radius and microstructure of the charged AdS BH, founding that the shadow can provide the dynamic characteristics of BHs \cite{24}. By constructing the connection between the shadow and Ruppeiner geometry, Wang $et~al.$ believed the BH shadow can provide information about the small/large phase transition of the charged AdS BH and obtained the normalized curvature scalar at the critical point has a constant value of $-3/2$ \cite{25}.

\par
Nevertheless, the above-mentioned connections between these BHs' shadow and thermodynamics are only for the singular spacetime. Penrose and Hawking's famous work showed that the existence of singularity is inevitable \cite{26}. In order to avoid singularity, Bardeen obtained the first regular BH solution in 1968 \cite{27}. Ay\'{o}n-Beato and Garc\'{\i}a pointed out the physical source of the regular BH is the nonlinear electrodynamics \cite{28}. We discussed the thermodynamic properties of the regular AdS BH and found their different features from the singular AdS BHs \cite{29,30}. Meanwhile, the observation feature of the shadows and rings for the regular BH are investigated \cite{31,32}. Based on these results, it is natural to ask whether the dependence of the BH shadow and thermodynamics may be structured in the regular spacetime and how the singularity affects this relation. This letter focuses on this issue.

\par
Taking the Bardeen-AdS BH as an example, we attempt to establish the relationship between the BH shadow and thermodynamic phase structure. The remainders of this letter are organized as follows. Sec.\ref{sec:2} calculates the shadow radius of the Bardeen-AdS BH and discusses the relationship between the shadow radius and the event horizon radius. Sec.\ref{sec:3} applies this relation to analyze the phase transition in the $T-r_{\rm s}$ plane. In Sec.\ref{sec:4}, we present the thermal profile of the Bardeen-AdS BH under the shadow context. We draw the conclusions and discussions in Sec.\ref{sec:5}. Here we use the units $G_{\rm N}=\hbar=\kappa_{\rm b}=c=1$.

\section{The shadow of the Bardeen-AdS BH}
\label{sec:2}
\par
The static spherically symmetric metric of the BH is given by \cite{28}
\begin{equation}
\label{2-1}
{\rm d}s^{2}=-f(r){\rm d}t^{2}+\frac{1}{u(r)}{\rm d}r^{2}+r^{2}{\rm d}\theta^{2}+r^{2}\sin^{2}\theta {\rm d}\phi^{2},
\end{equation}
where $f(r)$ and $u(r)$ are the function of the radius parameter $r$. The photons' motion around a BH satisfy the Euler-Lagrangian equation,
\begin{equation}
\label{2-2}
\frac{{\rm d}}{{\rm d}\lambda}\Bigg(\frac{\partial \mathcal{L}}{\partial \dot{x}^{\rm \mu}}\Bigg)=\frac{\partial \mathcal{L}}{\partial x^{\rm \mu}},
\end{equation}
in which $\lambda$ is affine parameter and $\dot{x}^{\rm \mu}$ is photon four-velocity. $\mathcal{L}$ is the Lagrangian density, it is
\begin{equation}
\label{2-3}
\mathcal{L}=-\frac{1}{2}g_{\mu \nu}\frac{{\rm d} x^{\mu}}{{\rm d} \lambda}\frac{{\rm d} x^{\nu}}{{\rm d} \lambda}=\frac{1}{2}\Big(f(r)\dot{t}^{2}-\frac{\dot{r}^{2}}{u(r)}-r^{2}(\dot{\theta}^{2}+\sin^{2}\theta \dot{\phi}^{2})\Big).
\end{equation}
In this letter, we consider that the photons move on the equatorial plane ($\theta=\pi/2$), so the metric does not depend explicitly on time $t$ and azimuthal angle $\phi$, the corresponding conserved constants as
\begin{equation}
\label{2-4}
E=-g_{\rm t \rm t}\frac{{\rm d} t}{{\rm d} \lambda}=f(r)\frac{{\rm d} t}{{\rm d} \lambda},~~~~ L=g_{\rm \phi \rm \phi}\frac{{\rm d} \phi}{{\rm d} \lambda}=r^{2} \frac{{\rm d} \phi}{{\rm d} \lambda},
\end{equation}
where $E$ and $L$ represent the energy and the angular momentum of the photons. According to the null geodesic $g_{\rm \mu \nu}\dot{x}^{\rm \mu}\dot{x}^{\rm \nu}=0$, the motion equation of photons is obtained,
\begin{eqnarray}
\label{2-5-1}
&&\frac{{\rm d} t}{{\rm d} \lambda}= \frac{E}{f(r)},\\
\label{2-5-2}
&&\frac{{\rm d}\phi}{{\rm d} \lambda}= \frac{L}{r^{2}},\\
\label{2-5-3}
&&\frac{{\rm d} r}{{\rm d} \lambda}= \pm \frac{\sqrt{u(r)E^{2}r^{4}-u(r)L^{2}r^{2}f(r)}}{r^{2}\sqrt{f(r)}},
\end{eqnarray}
where the symbol ``$\pm$" indicates the counterclockwise ($-$) and clockwise ($+$) direction for the motion of photons. These three equations report a complete description of the photon dynamics around the BH at which the effective potential can be written as
\begin{equation}
\label{2-6}
\Bigg(\frac{{\rm d} r}{{\rm d} \lambda}\Bigg)^{2} + \mathcal{V}_{\rm eff}(r) = 0,
\end{equation}
where
\begin{equation}
\label{2-7}
\mathcal{V}_{\rm eff}(r) = u(r)\Bigg(\frac{L^2}{r^2}-\frac{E^{2}}{f(r)}\Bigg).
\end{equation}
The critical photon ring orbit satisfies the effective potential critical conditions
\begin{equation}
\label{2-8}
\mathcal{V}_{\rm eff}(r)=0,~~~~\mathcal{V}'_{\rm eff}(r)=0,~~~~\mathcal{V}''_{\rm eff}(r)>0.
\end{equation}
Substituting Eq.(\ref{2-7}) into Eq.(\ref{2-8}), one can obtain $\frac{L}{E}= \frac{r_{\rm pp}}{\sqrt{f(r_{\rm pp})}}$, where $r_{\rm pp}$ is the radius of the photon ring. Based on Eqs.(\ref{2-5-2}) and (\ref{2-5-3}), we have
\begin{equation}
\label{2-9}
\frac{{\rm d} r}{{\rm d} \phi} = \Bigg(\frac{{\rm d} r}{{\rm d} \lambda}\Bigg) {\Bigg /} \Bigg(\frac{{\rm d}\phi}{{\rm d} \lambda}\Bigg) = \pm r \sqrt{\frac{r^{2}E^{2}u(r)}{L^{2}f(r)}-u(r)}.
\end{equation}
Considering that the turning point of photon orbits, satisfying $\frac{{\rm d} r}{{\rm d} \phi}{\big |}_{r=\chi}=0$, the Eq.(\ref{2-9}) can be rewritten as
\begin{equation}
\label{2-10}
\frac{{\rm d} r}{{\rm d} \phi} = \pm r \sqrt{\frac{r^{2}f(\chi)^{2}u(r)}{\chi^{2}f(r)}-u(r)}.
\end{equation}
Following Ref. \cite{22,23,24,25}, the light ray send from a static observer at position $r_{\rm O}$ transmits into the past with an angle $\beta$ relative to the radial direction, that is
\begin{equation}
\label{2-11}
\cot \beta = \frac{\sqrt{g_{\rm rr}}}{\sqrt{g_{\rm \phi \phi}}}{\Bigg |}_{r=r_{\rm O}} = \pm \sqrt{\frac{r_{\rm O}^{2}f(\chi)}{\chi^{2}f(r_{\rm O})}-1}.
\end{equation}
With the elementary trigonometry, one can get
\begin{equation}
\label{2-12}
\sin^{2} \beta = \frac{\chi^{2}f(r_{\rm O})}{r_{\rm O}^{2}f(\chi)},
\end{equation}
the shadow radius of the BH observed by a static observer at $r_{\rm O}$ can be written as
\begin{equation}
\label{2-13}
r_{\rm ss}= r_{\rm O} \sin \beta = \chi\sqrt{\frac{f(r_{\rm O})}{f(\chi)}}{\Bigg |}_{\chi \rightarrow r_{\rm pp}}.
\end{equation}
For the regular Bardeen-AdS BH, the metric potential can be expressed by \cite{33}
\begin{equation}
\label{2-14}
f(r)= 1 + \frac{8\pi P r^{2}}{3}-\frac{2 M r^{2}}{(r^{2}+g^{2})^{3/2}},
\end{equation}
where $M$ is the BH mass and $g$ is the BH magnetic charge. The radius of the event horizon $r_{\rm h}$ is the largest root of $f(r_{\rm h})=0$. The BH mass reads as
\begin{equation}
\label{2-15}
M=\frac{(3+8 P \pi r_{\rm h}^{2})(r_{\rm h}^{2}+g^{2})^{{3}/{2}}}{6 r_{\rm h}^{2}}.
\end{equation}
Based on the first law of BH thermodynamics, the Bardeen-AdS BH temperature is
\begin{equation}
\label{2-16}
T = \frac{r_{\rm h}^{2}+8 \pi P r_{\rm h}^{4}-2g^{2}}{4 \pi r_{\rm h} (r_{\rm h}^{2}+g^{2})}.
\end{equation}
The state equation can be expressed as
\begin{equation}
\label{2-17}
P = \frac{T}{2 r_{\rm \rm h}}-\frac{1}{8 \pi {r_{\rm h}}^{3}}+\frac{g^2 T}{2r_{\rm h}^{3}}+\frac{g^{2}}{4 \pi {r_{\rm h}}^{4}}.
\end{equation}
According to Eq.(\ref{2-8}), the photon circular orbit radius of the Bardeen-AdS BH is
\begin{equation}
\label{2-18}
r_{\rm p} = \frac{1}{2}\sqrt{3 M + \sqrt{9 M^{2} - 10 g^{2}}}.
\end{equation}
Utilizing Eq.(\ref{2-13}), the radius of the Bardeen-AdS BH shadow can be written as
\begin{equation}
\label{2-19}
r_{\rm s} = r_{\rm p} \sqrt{\frac{f(r_{\rm O})}{f(r_{\rm p})}},
\end{equation}
Thinking of Eqs.(\ref{2-14}), (\ref{2-18}), and (\ref{2-19}), we have the specific expression of the Bardeen-AdS BH shadow radius related to the horizon radius, i.e. Eq.(\ref{2-20}).
\begin{equation}
\label{2-20}
r_{\rm s}=\frac{1}{2}\sqrt{\frac{3 f(r_{\rm O}) A^{2}B}{15(r_{\rm h}^{2}+g^{2})^{{3}/{2}}(3+8 P \pi r_{\rm h}^{2})+3B+2 \pi P A^{2}B}},
\end{equation}
where $A \equiv \frac{(r_{\rm h}^{2}+g^{2})^{{3}/{2}}(3+8 P \pi r_{\rm h}^{2})}{2r_{\rm h}^{2}}+ \Big(\frac{(r_{\rm h}^{2}+g^{2})^{3} (3+8 \pi P r_{\rm h}^{2})^{2} }{4r_{\rm h}^{4}}-10g^{2}\Big)^{-1/2}$, $B \equiv 6(r_{\rm h}^{2}+g^{2})^{3/2}(3+8 \pi P r_{\rm h}^{2})+3r_{\rm h}^{2}\Big(\frac{(r_{\rm h}^{2}+g^{2})^{3} (3+8 \pi P r_{\rm h}^{2})^{2} }{4r_{\rm h}^{4}}-10g^{2}\Big)^{-1/2}$.

\par
Based on the constraint, a static observer at spatial infinity have $f(r_{\rm O})=1$ for $r_{\rm O}=100$ \cite{22}, and when the magnetic charge $g \rightarrow 0$, the Bardeen-AdS BH will degenerate into the Schwarzschild-AdS BH. Fig.1 shows the radius of the Bardeen-AdS BH shadow $r_{\rm s}$ as a function of the event horizon radius $r_{\rm h}$ under several representative values of magnetic charge. One can see that $r_{\rm s}$ and $r_{\rm h}$ display a positive correlation, meaning that the BH temperature can be rewritten as a function of shadow radius in the regular spacetime. Meanwhile, the trend of $r_{\rm s}$ with the increase of $r_{\rm h}$ is gradually flat. As the magnetic charge increases, the radius of the event horizon increases, and the corresponding shadow radius gets larger. For the magnetic charge equal to zero, the regular Bardeen-AdS BH is replaced by the singular Schwarzschild-AdS BH, so the Eq.(\ref{2-20}) can be used to describe the relationship between the shadow radius and the event horizon radius of the Schwarzschild-AdS BH ($g=0$). The left panels of Fig.1 reports the shadow radius of the Schwarzschild-AdS BH as a function of the event horizon radius. It is observe that these two radii still show positive correlation characteristics, while the slope of function curves are greater than regular AdS BH. As a result, we believe that the relationship between shadow and BH temperature can be constructed in regular spacetime.
\begin{center}
  \includegraphics[width=5cm,height=4.2cm]{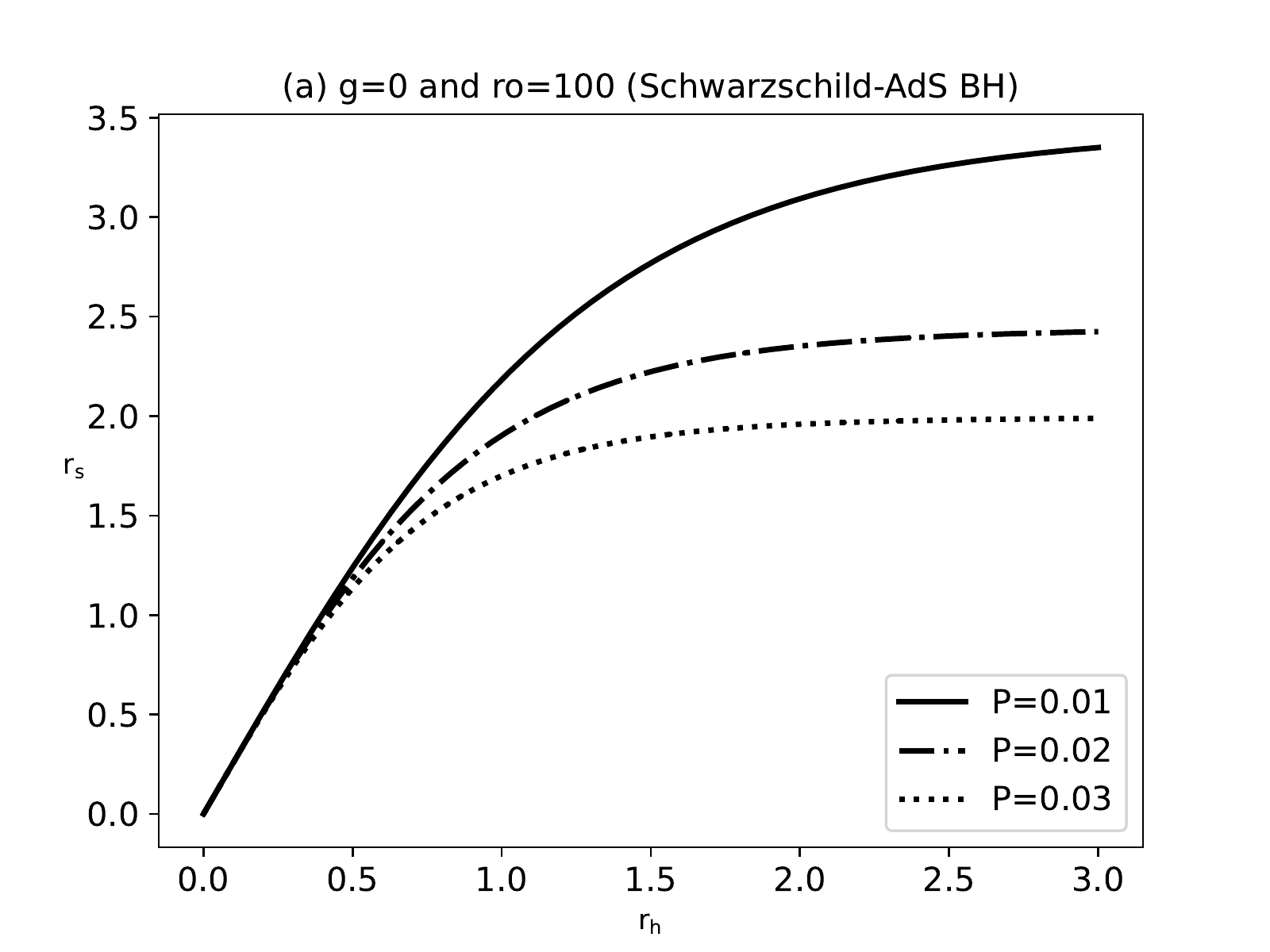}
  \includegraphics[width=5cm,height=4.2cm]{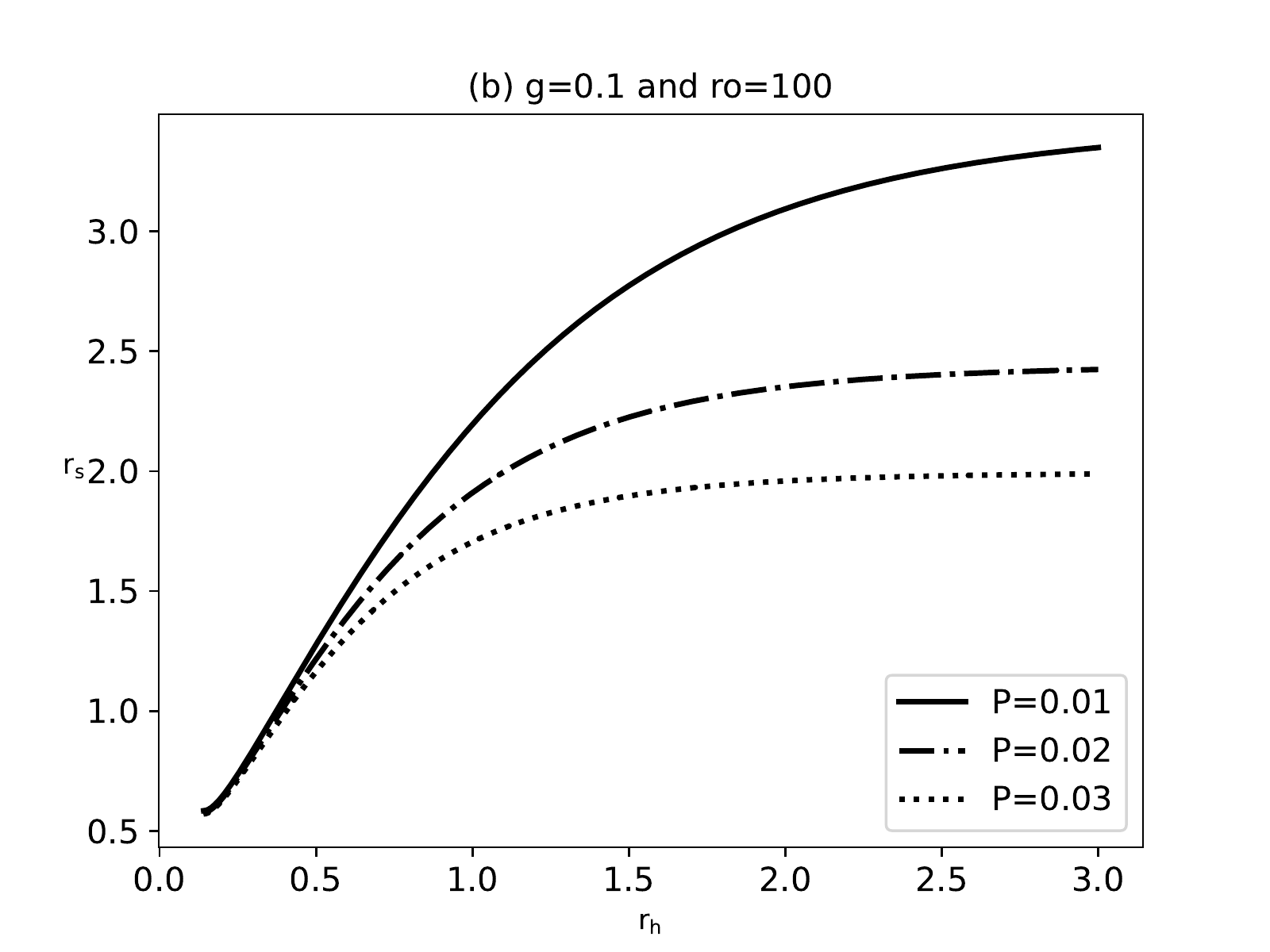}
  \includegraphics[width=5cm,height=4.2cm]{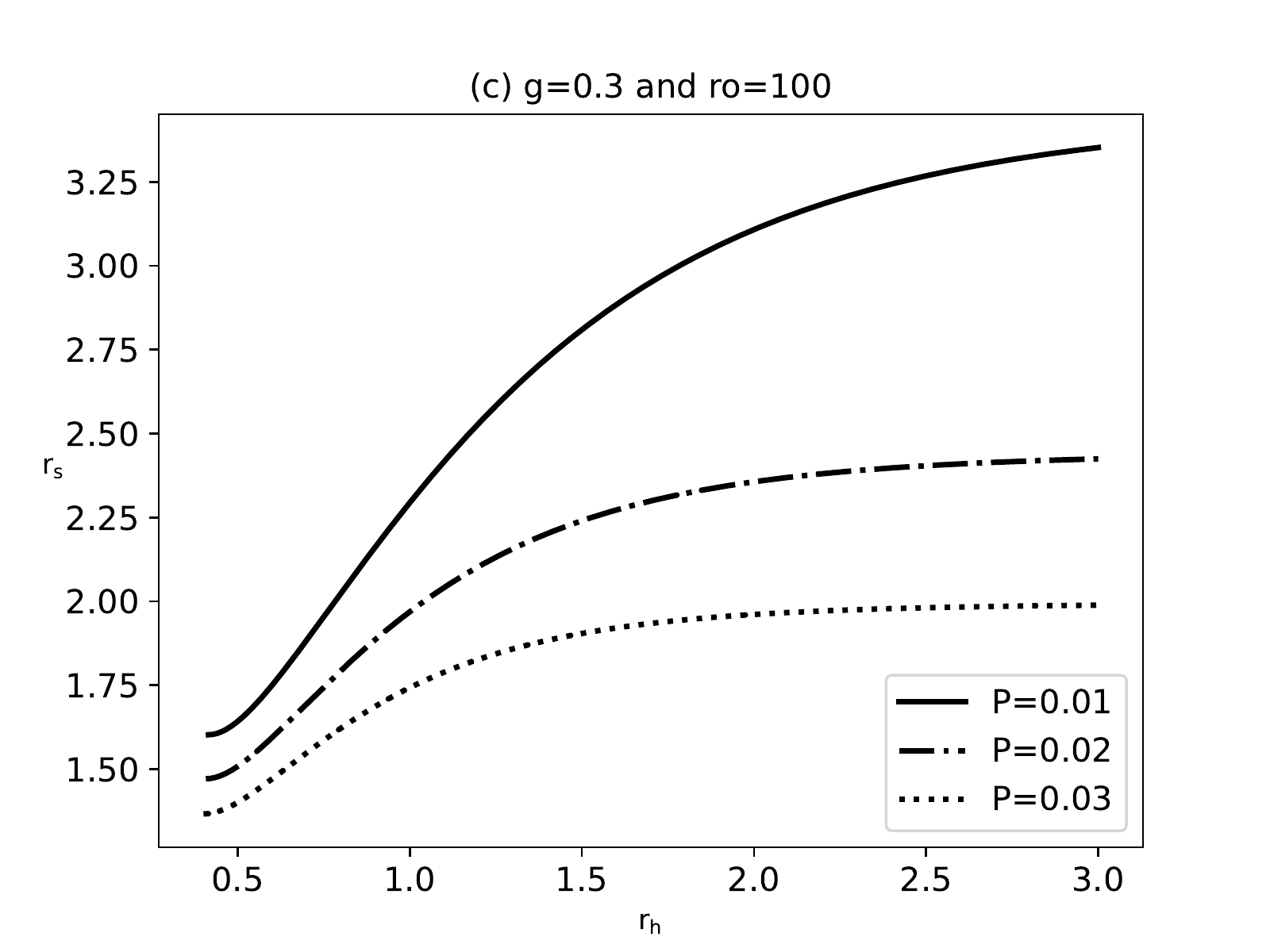}
\parbox[c]{15.0cm}{\footnotesize{\bf Fig~1.}
The variation of shadow radius $r_{\rm s}$ in terms of the event horizon radius $r_{\rm h}$ for the Bardeen-AdS BH. {\em Panel (a)}-- magnetic charge $g=0$ and $r_{\rm O}=100$ (Schwarzschild-AdS BH), {\em Panel (b)}-- magnetic charge $g=0.1$ and $r_{\rm O}=100$, {\em Panel (c)}-- magnetic charge $g=0.3$ and $r_{\rm O}=100$. The solid lines, segment point lines and dotted lines correspond to $P=0.001$, $P=0.002$ and $P=0.003$, respectively.}
\label{fig1}
\end{center}

\section{Phase transition of the Bardeen-AdS BH using shadow analysis}
\label{sec:3}
This section investigates the phase transition of the Bardeen-AdS BH from the shadow perspective. According to the state equation (\ref{2-17}) and the critical condition $({\partial P}/{\partial r_{\rm h}})=0=({{\partial}^{2}P}/{\partial r_{\rm h}^{2}})$, the critical thermodynamic quantities of the Bardeen-AdS BH are obtained,
\begin{eqnarray}
\label{3-1-1}
&&P_{\rm c}=\frac{219-13\sqrt{273}}{1152 \pi g^{2}} \simeq {0.00116}{g^{-2}},\\
\label{3-1-2}
&&r_{\rm c}=\frac{\sqrt{15g^{2}+\sqrt{273}g^{2}}}{\sqrt{2}} \simeq 3.97006g,\\
\label{3-1-3}
&&T_{\rm c}=\frac{\sqrt{2(15+\sqrt{273})}}{(51 \pi + 3\sqrt{273}\pi)g} \simeq {0.02513}{g^{-1}}.
\end{eqnarray}
Based on Eqs.(\ref{2-20})-(\ref{3-1-2}), the critical shadow radius of the Bardeen-AdS BH is $r_{\rm sc} \simeq 7.91886g$. It is shown that the critical shadow radius is about twice larger than the standard critical radius. Considering the temperature expression Eq.(\ref{2-16}), the BH temperature as a function of the $r_{\rm h}$ for a fixed value of the magnetic charge is shown in the left panels of Fig.2. We can see that the function curves show different characteristics for different pressures. Above the critical isobaric ($P>P_{\rm c}$), the curve does not have inflection point. The temperature is a monotonically increasing function of radius, indicating that the BH is in the supercritical phase. The critical isobaric ($P=P_{\rm c}$) has an inflection point, the BH is thermodynamically unstable, corresponding to the critical temperature. Two-phase transition branches exist below the critical pressure ($P<P_{\rm c}$), one is in the small radius region, corresponding to the fluid phase in the vdW system, and the other is in the large radius region, corresponding to the gas phase.
\begin{center}
  \includegraphics[width=5cm,height=4.2cm]{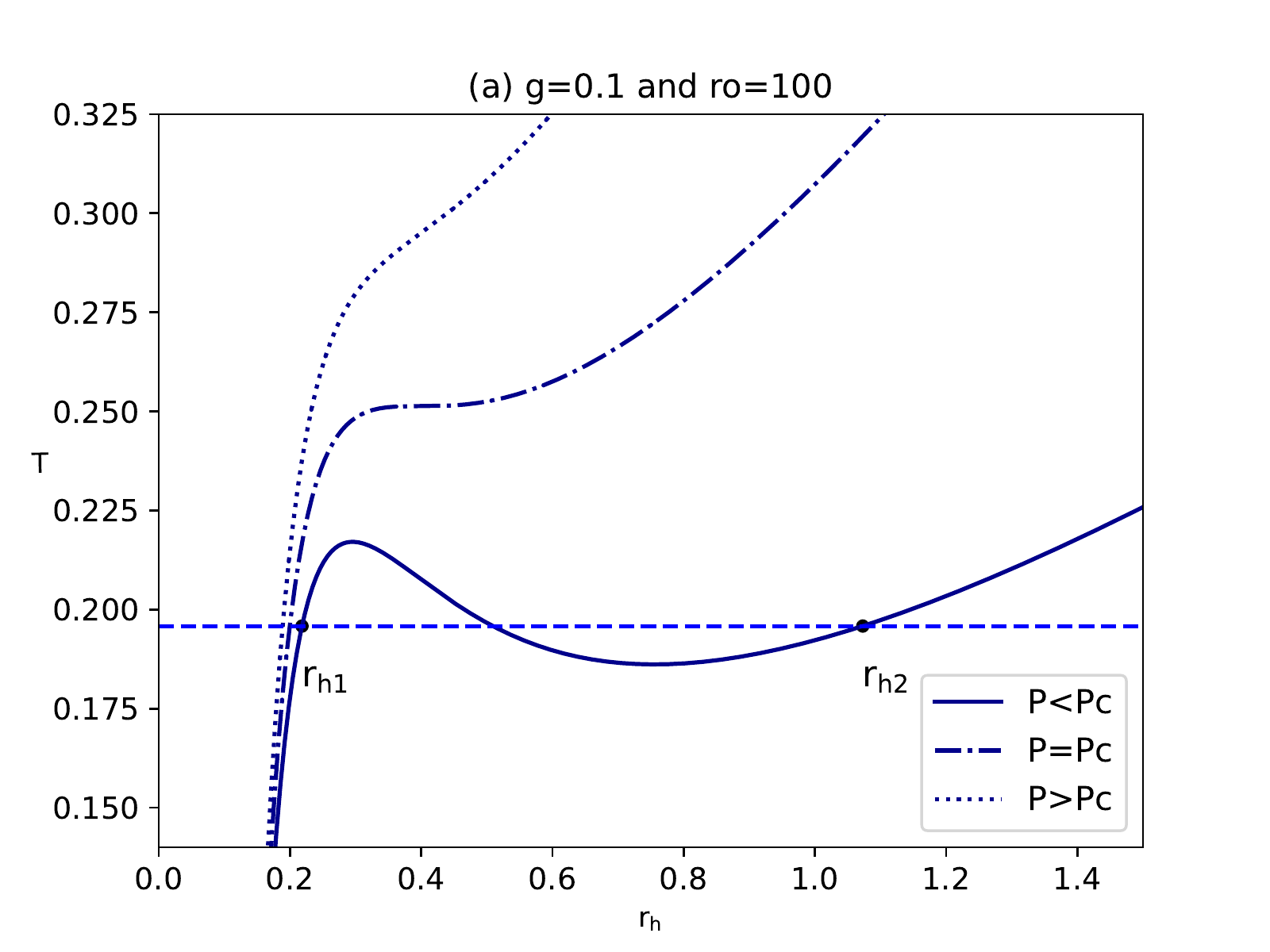}
  \includegraphics[width=5cm,height=4.2cm]{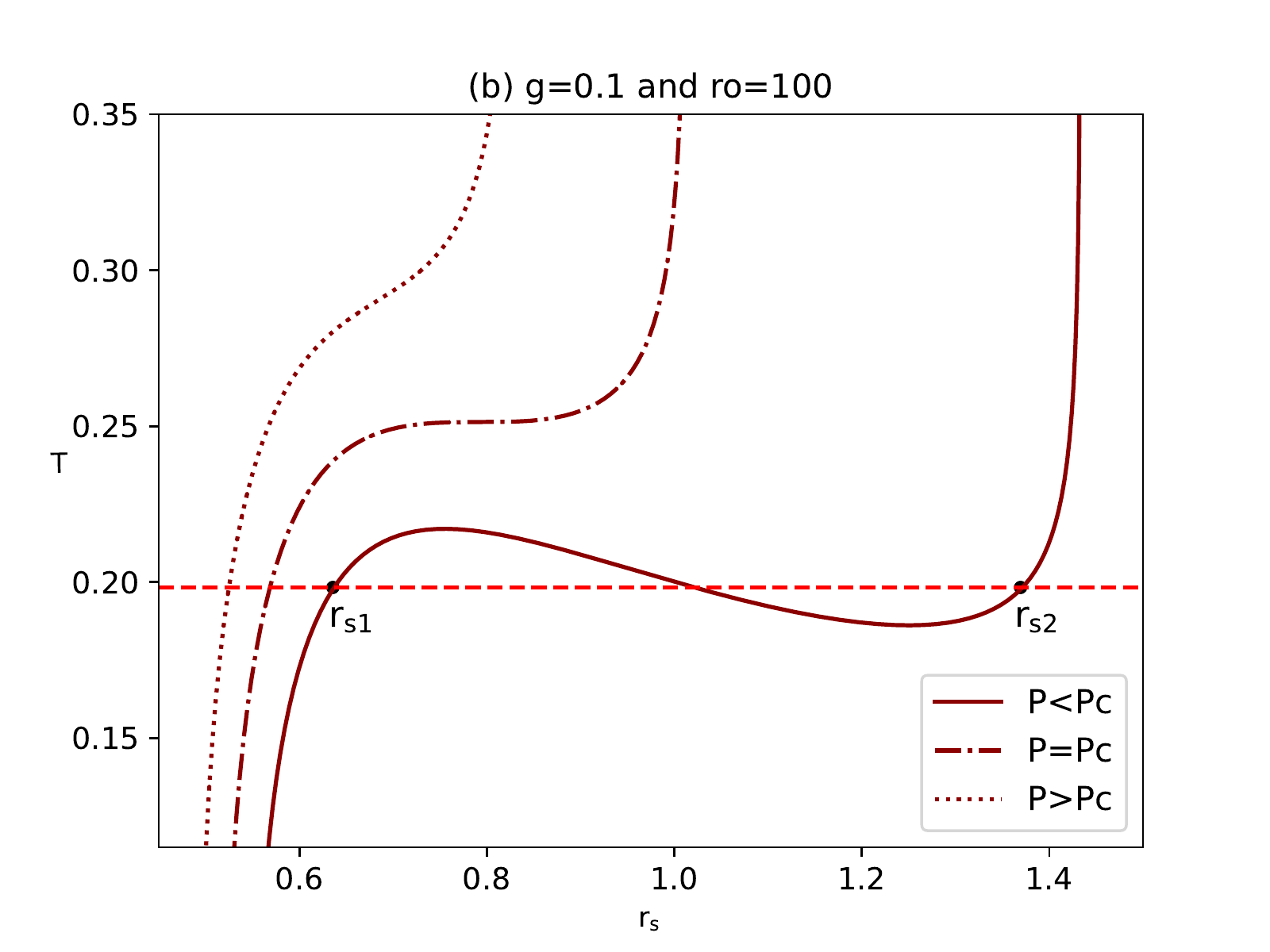}
  \includegraphics[width=5cm,height=4.2cm]{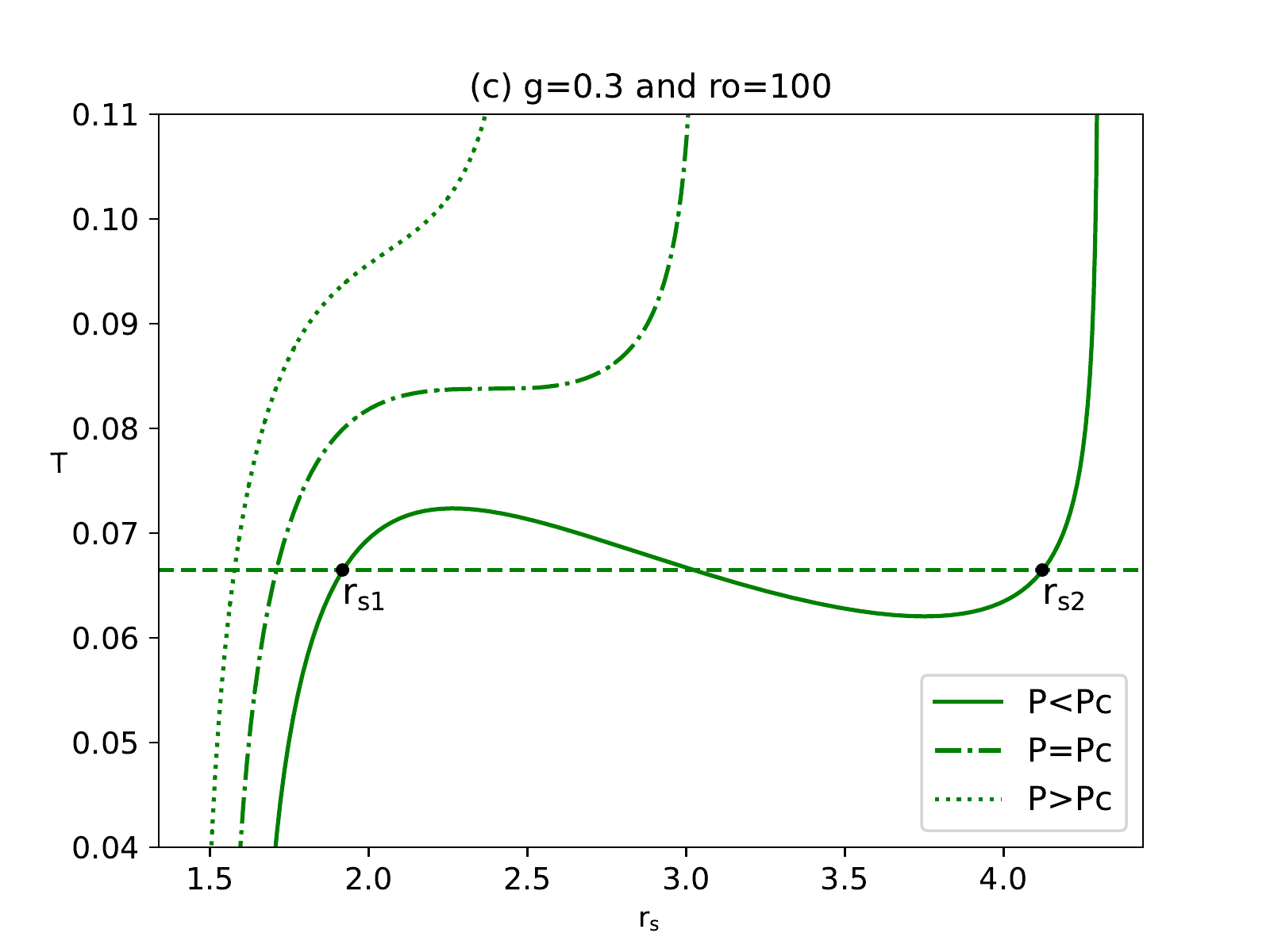}
\parbox[c]{15.0cm}{\footnotesize{\bf Fig~2.}
{\em Panel (a)}-- temperature as a function of $r_{\rm h}$ with $g=0.1$, {\em Panel (b)}-- temperature as a function of $r_{\rm s}$ with $g=0.1$, {\em Panel (c)}-- temperature as a function of $r_{\rm s}$ with $g=0.3$. A static observer at $r_{\rm O}=100$.}
\label{fig2}
\end{center}

\par
The middle and right panels of Fig.2 present the BH temperature as a function of the shadow radius $r_{\rm s}$ with different pressure under several representative values of magnetic charge. One can observe that the shadow radius can replace the event horizon radius to present the Bardeen-AdS BH phase transition process. The $r_{\rm s}$ less than $r_{\rm s1}$ corresponds to a stable small BH, and the $r_{\rm s}$ great than $r_{\rm s2}$ corresponds to a stable large BH. The unstable intermediate branch appears in $(r_{\rm s1}$, $r_{\rm s2})$. A larger magnetic charge leads to a weaker phase transition temperature and the corresponding shadow radius get larger. Meanwhile, Maxwell's equal area is constructed in the $T-r_{\rm h}$ plane and $T-r_{\rm s}$ plane, i.e.
\begin{eqnarray}
\label{3-2-1}
&&T_{\rm h0}(r_{\rm h2}-r_{\rm h1}) = {\int_{r_{\rm h2}}^{r_{\rm h1}} T {\rm d}r_{\rm h}},\\
\label{3-2-2}
&&T_{\rm s0}(r_{\rm s2}-r_{\rm s1}) = {\int_{r_{\rm s2}}^{r_{\rm s1}} T {\rm d}r_{\rm s}}.
\end{eqnarray}

\par
We obtained that the equal area law can also be established with shadow radius in the regular spacetime, which implies that the shadow radius may serve as a probe for the phase structure in the regular spacetime. Note that $T_{\rm h0}$ and $T_{\rm s0}$ are not completely equivalent because the position of the static observer is relatively remote. In transferring the phase transition results to the characterization of shadow radius, the temperature increases slightly with the Hawking radiation.

\par
On the other hand, the phase transition grade can be determined by heat capacity. The heat capacity mutation and specific heat diverge represent the second-order phase transition at the critical point. The heat capacity of the Bardeen-AdS BH at constant pressure can be written as
\begin{equation}
\label{3-3}
C_{\rm P}=T\Bigg(\frac{dS}{dT}\Bigg)_{\rm P,g}=\frac{2\pi r_{\rm h}^2 (g^{2}+r_{\rm h}^2)(8\pi P r_{\rm h}^4 -2g^{2}+r_{\rm h}^2)}{8\pi P r_{\rm h}^6+8g^{4}+4g^{2}r_{\rm h}^2-r_{\rm h}^4}.
\end{equation}

\par
Fig.3 shows the heat capacity as a function of the event horizon radius and the shadow radius for the Bardeen-AdS BH. It is found that the specific heat shows infinite divergence at the critical point, which is a strong signal for the beginning of the higher-order phase transition. By exploring the relationship between $C_{\rm P}$ and $r_{\rm s}$, the shadow radius can also reveal the BH phase transition grade.
\begin{center}
  \includegraphics[width=5cm,height=4.2cm]{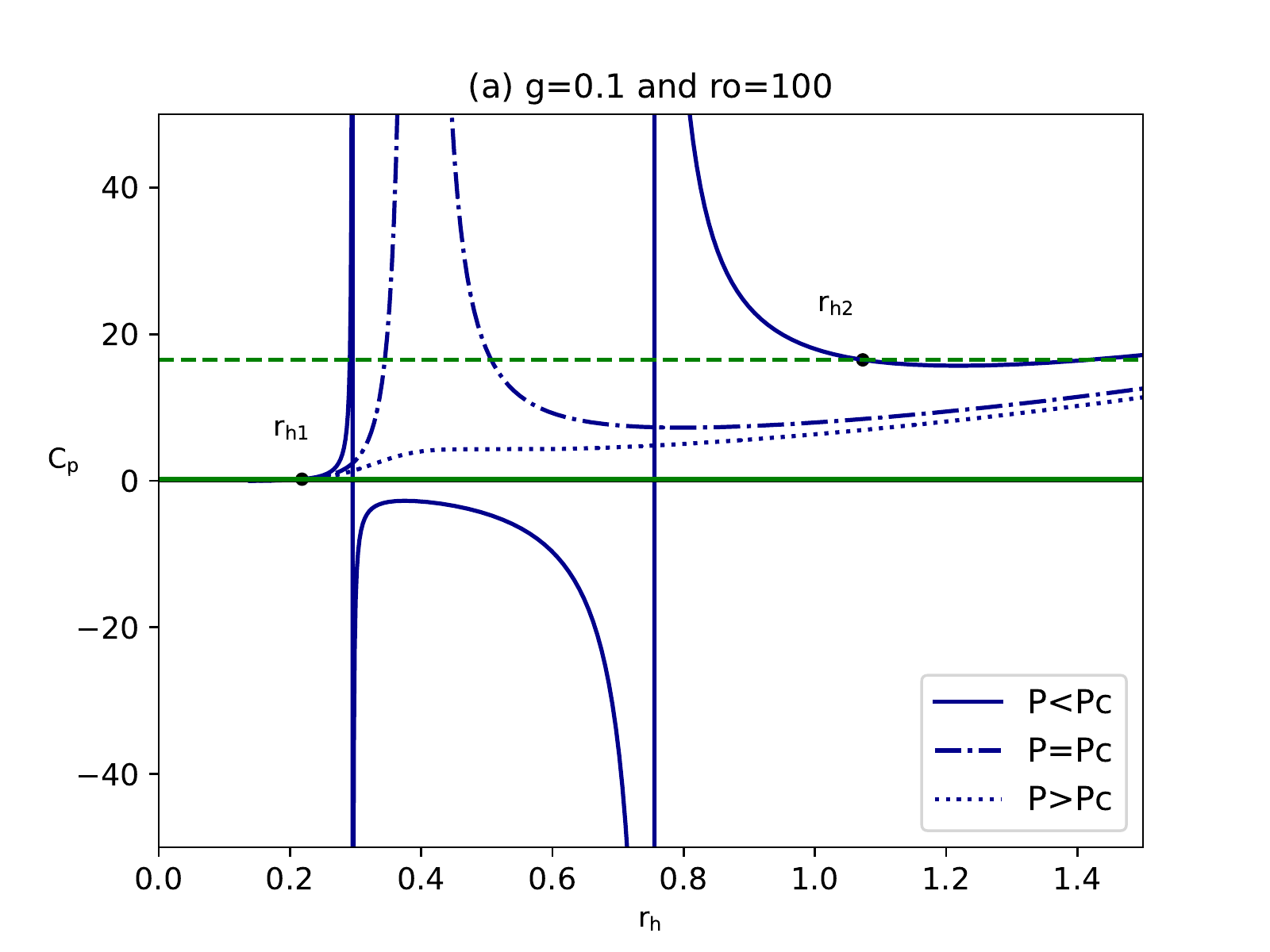}
  \includegraphics[width=5cm,height=4.2cm]{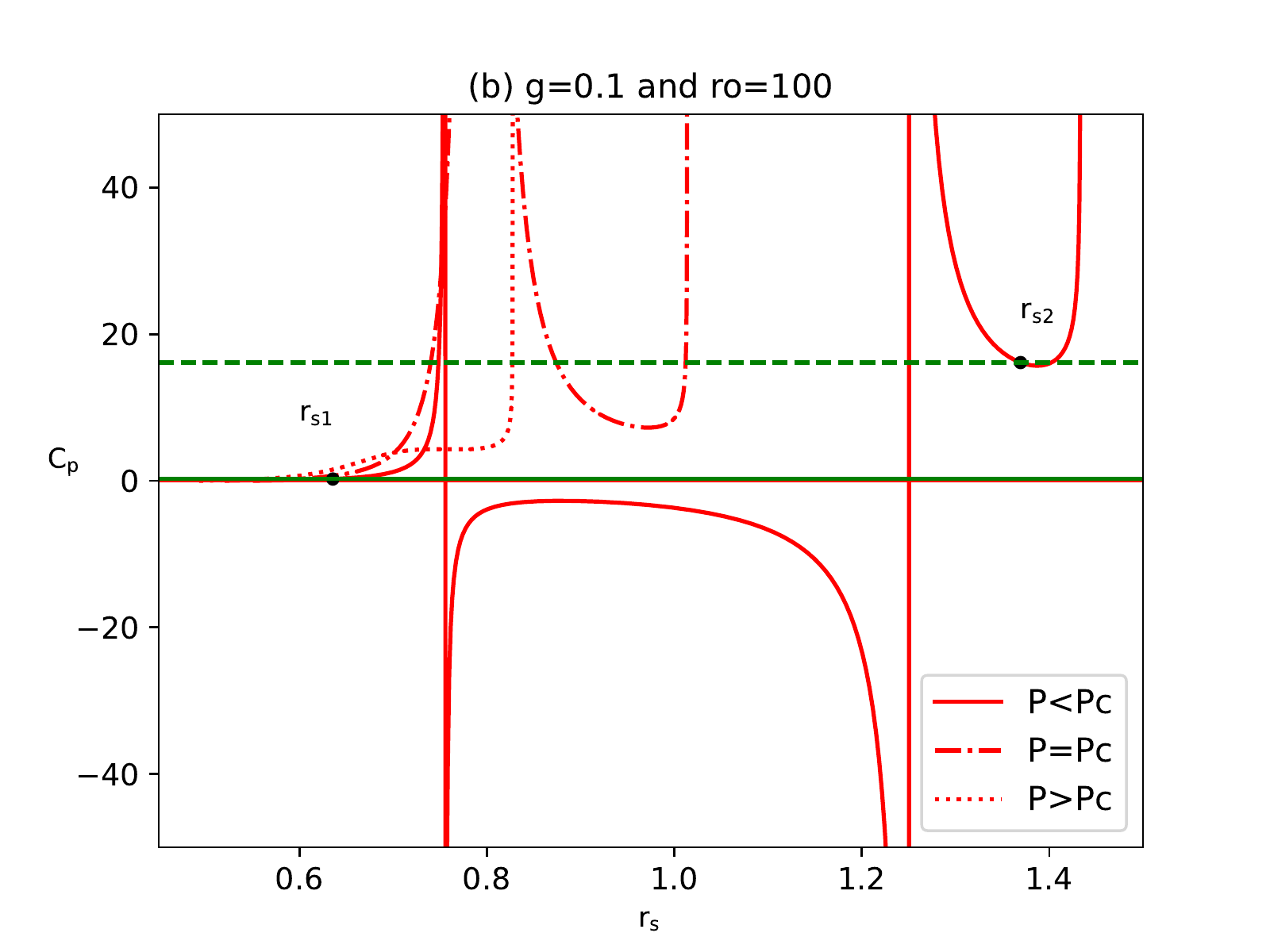}
  \includegraphics[width=5cm,height=4.2cm]{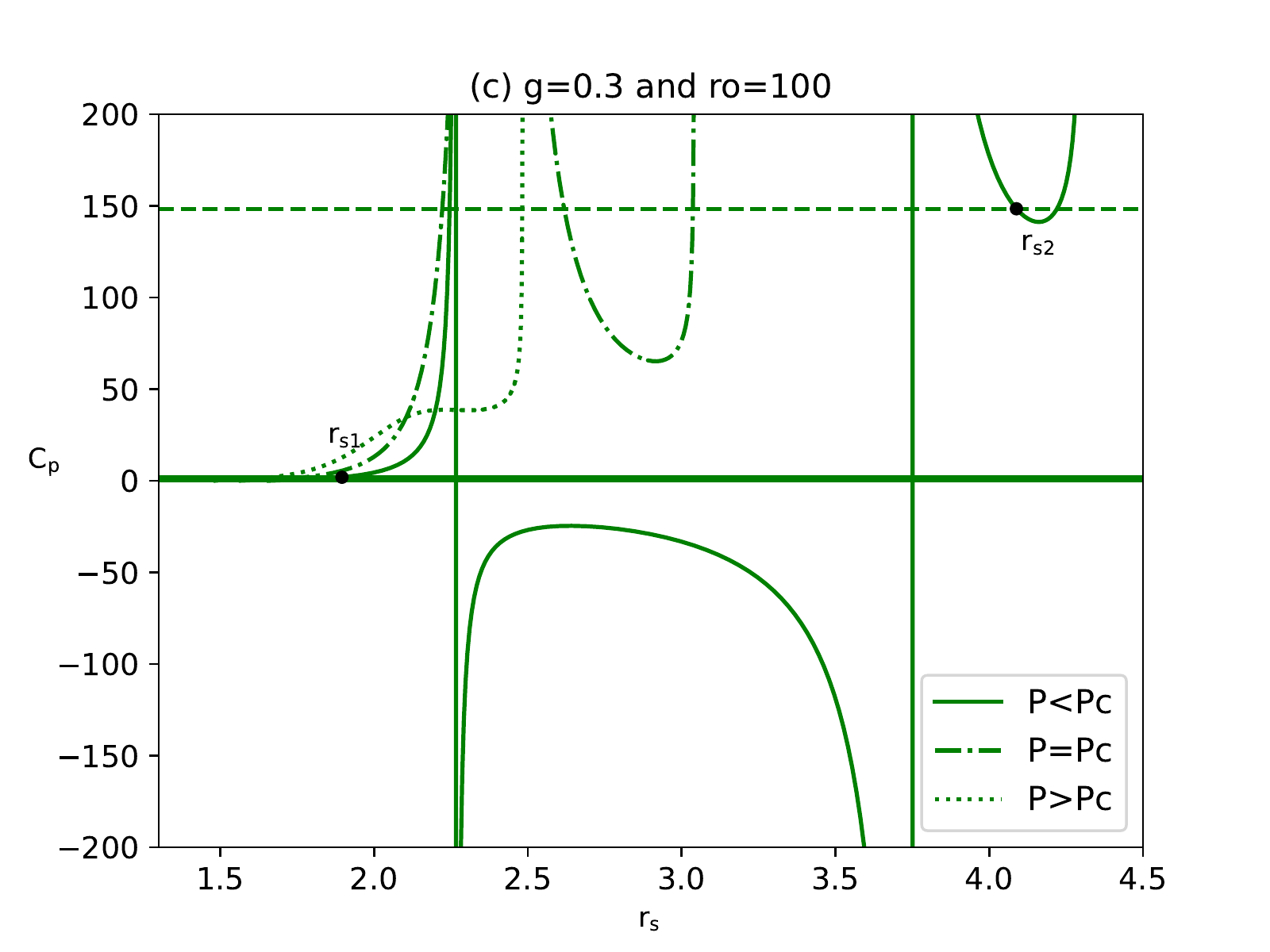}
\parbox[c]{15.0cm}{\footnotesize{\bf Fig~3.}
{\em Panel (a)}-- heat capacity as a function of $r_{\rm h}$ with $g=0.1$, {\em Panel (b)}-- heat capacity as a function of $r_{\rm s}$ with $g=0.1$. {\em Panel (c)}-- heat capacity as a function of $r_{\rm s}$ with $g=0.3$. A static observer at $r_{\rm O}=100$.}
\label{fig3}
\end{center}

\section{Thermal profile of the Bardeen-AdS BH}
\label{sec:4}
\par
We establish a thermal profile in this section to more intuitively reports the relationship between the BH phase structure and shadow in the regular spacetime. According to Ref. \cite{34}, the shadow boundary curve at the celestial coordinate reads as
\begin{eqnarray}
\label{3-2-1}
&& x = \lim\limits_{r \rightarrow \infty}\Big(-r^{2}\sin\theta_{0}\frac{{\rm d}\phi}{{\rm d}r}\Big)_{\rm \theta_{\rm 0} \rightarrow \frac{\pi}{2}},\\
\label{3-2-2}
&& y = \lim\limits_{r \rightarrow \infty}\Big(r^{2}\frac{{\rm d}\theta}{{\rm d}r}\Big)_{\rm \theta_{\rm 0} \rightarrow \frac{\pi}{2}}.
\end{eqnarray}

\par
Fig.4 illustrates the shadow contour for a static observer. It is found that the size of the BH shadow depends on the pressure. The point line represents $P>P_{\rm c}$, the shadow corresponding to the BH at the supercritical phase. The segment point line corresponds to the $P=P_{\rm c}$, we can find that the shadow radius more significant than the $P>P_{\rm c}$ situation. The dotted line only supports the $P<P_{\rm c}$ in which the shadow radius is in the large radius region. Additionally, we also observe that the radius of the BH shadow is expand outward the BH by increasing the magnetic charge.
\begin{center}
\includegraphics[width=5.5cm,height=5.5cm]{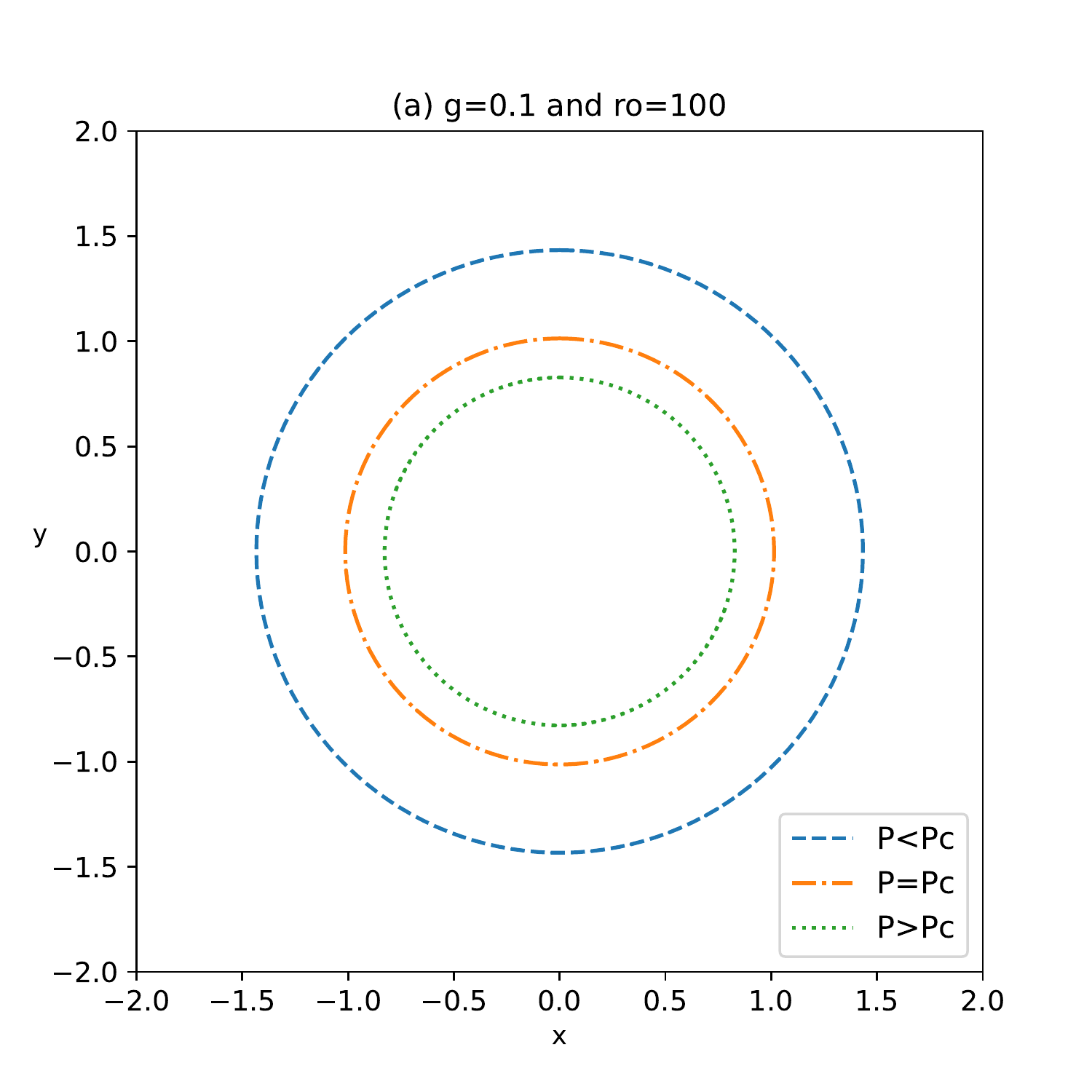}
\includegraphics[width=5.5cm,height=5.5cm]{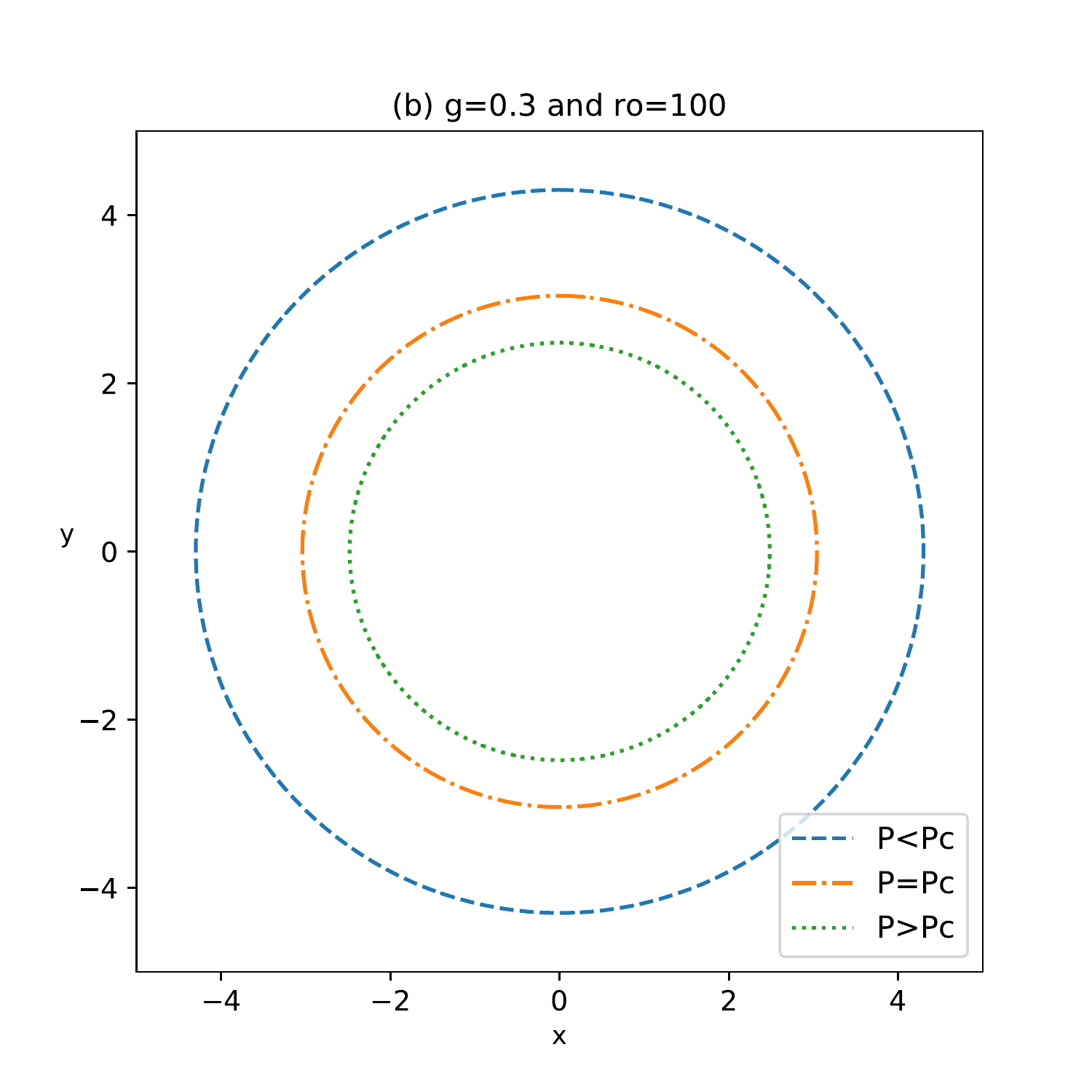}
\parbox[c]{15.0cm}{\footnotesize{\bf Fig~4.}
Shadow cast of the Bardeen-AdS BH. {\em Panel (a)}-- magnetic charge $g=0.1$, {\em Panel (b)}-- magnetic charge $g=0.3$. Here, the BH mass is $M=60$ and $r_{\rm O}=100$.}
\label{fig4}
\end{center}

\par
Consider overlaying Figs.2 and 4, the BH thermal profile is built by utilizing the temperature-shadow radius function. Under several representative values of magnetic charge, Fig.5 present three different scenarios: $P>P_{\rm c}$, $P=P_{\rm c}$, and $P<P_{\rm c}$. It is found that the change of temperature with $r_{\rm s}$ in the celestial coordinate is consistent with the previous analysis results. As the pressure decreases, the radius of the thermal profile increases, and the corresponding BH temperature gets smaller. The left panels of Fig.5 represent $P>P_{\rm c}$, showing the temperature increases gradually from the center of the shadow to the boundary in this case. It corresponds to the dotted lines in Fig.2. The middle panels of Fig.5 correspond to the $P=P_{\rm c}$ situation, the BH is thermodynamically unstable, and the temperature remains constant in the critical region, corresponding to the segment point lines in Fig.2. The right panels of Fig.5 have support in the $P<P_{\rm c}$. In this sense, the temperature shows an N-type change trend. We further refine the region from $r_{\rm s1}$ to $r_{\rm s2}$. Fig.6 show that the temperature variation law is: increasing $\rightarrow$ decreasing $\rightarrow$ increasing (N-type), which corresponds to the description of solid lines in Fig.2. Note that the increase of the BH magnetic charge $g$ leads to the decrease of the phase transition temperature, and there is a large unstable phase transition region. Our results suggest that the phase transition process of the Bardeen-AdS BH can be presented entirely in the thermal profile.
\begin{center}
  \includegraphics[width=5cm,height=4cm]{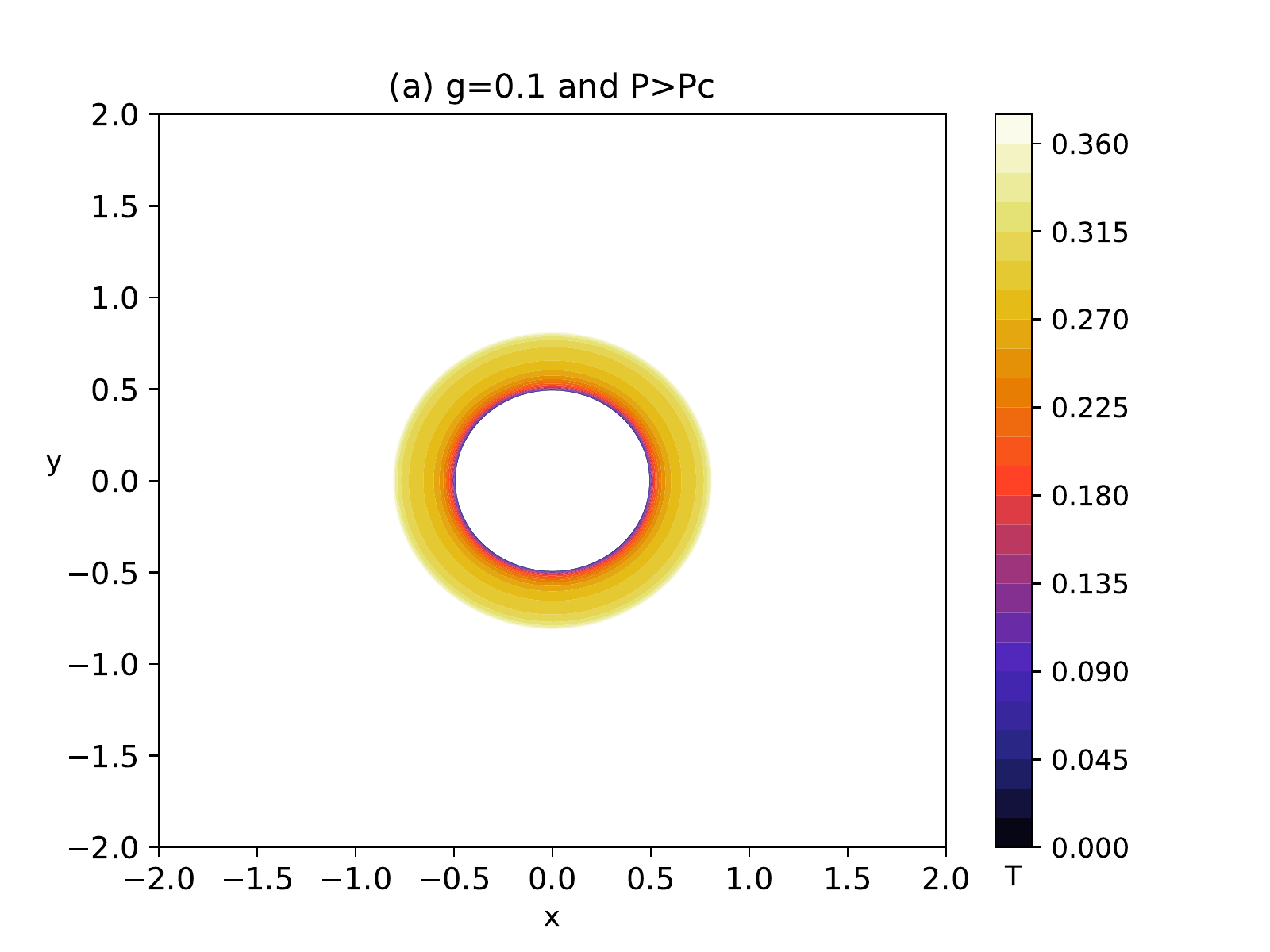}
  \includegraphics[width=5cm,height=4cm]{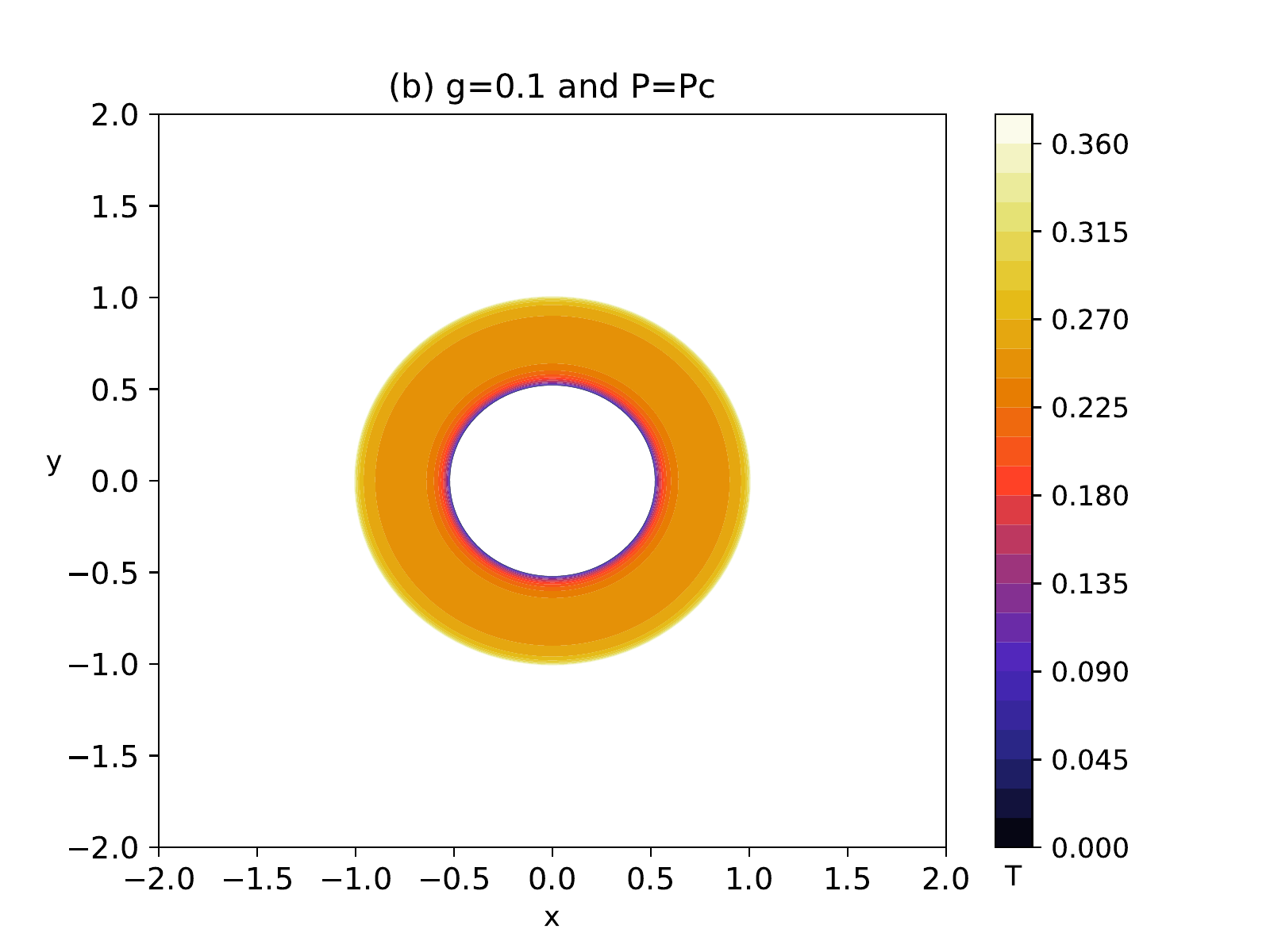}
  \includegraphics[width=5cm,height=4cm]{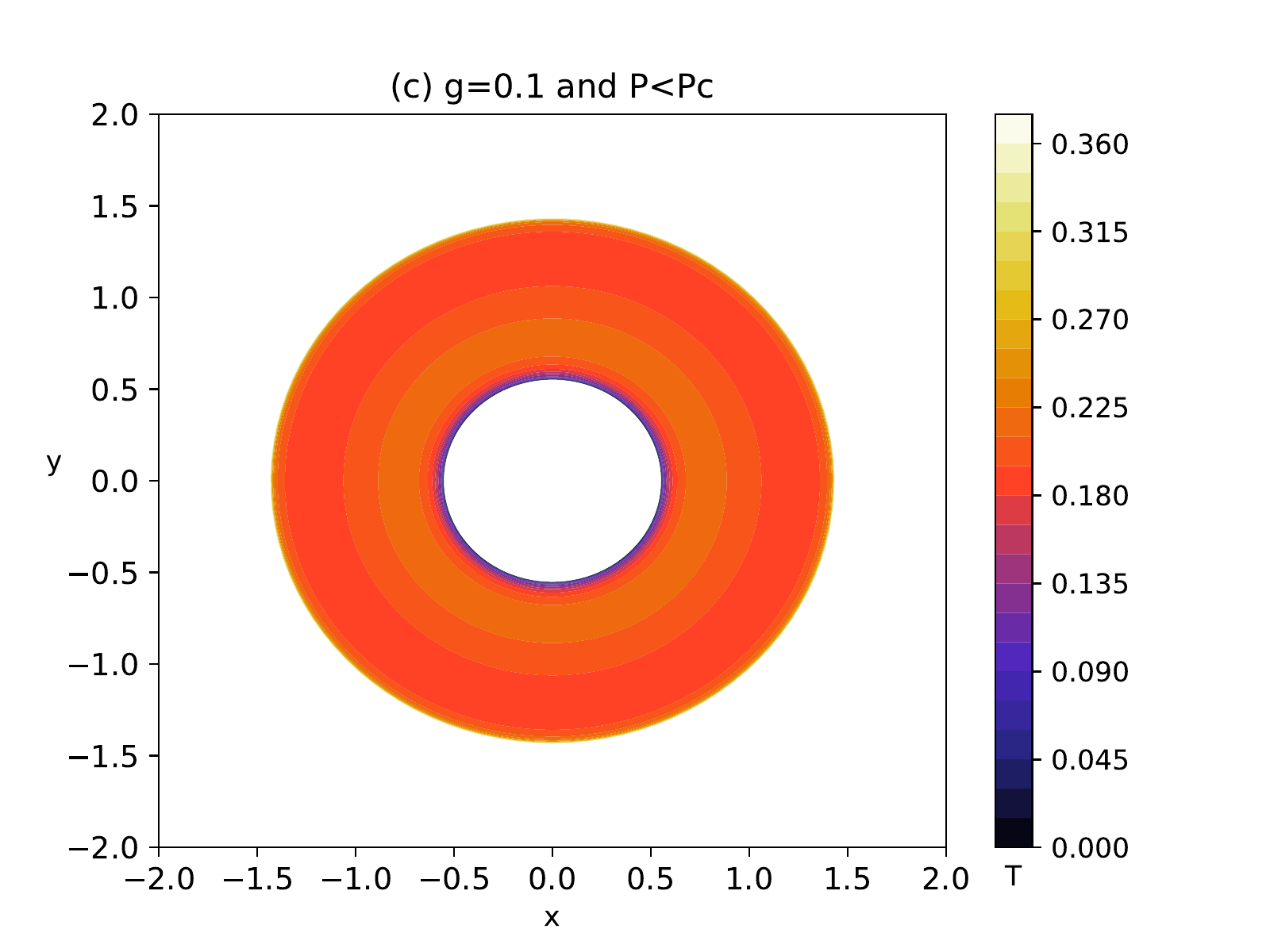}
  \includegraphics[width=5cm,height=4cm]{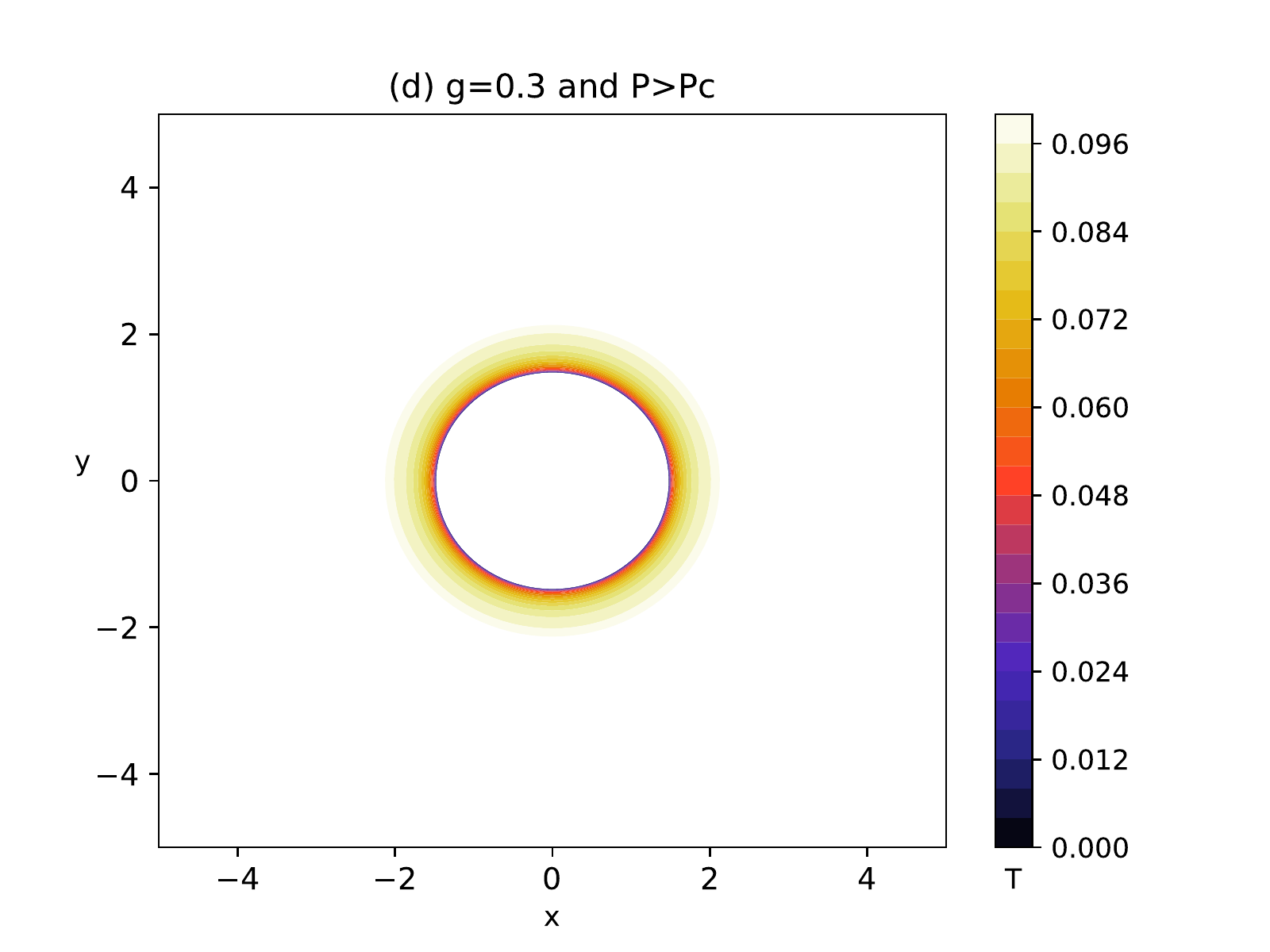}
  \includegraphics[width=5cm,height=4cm]{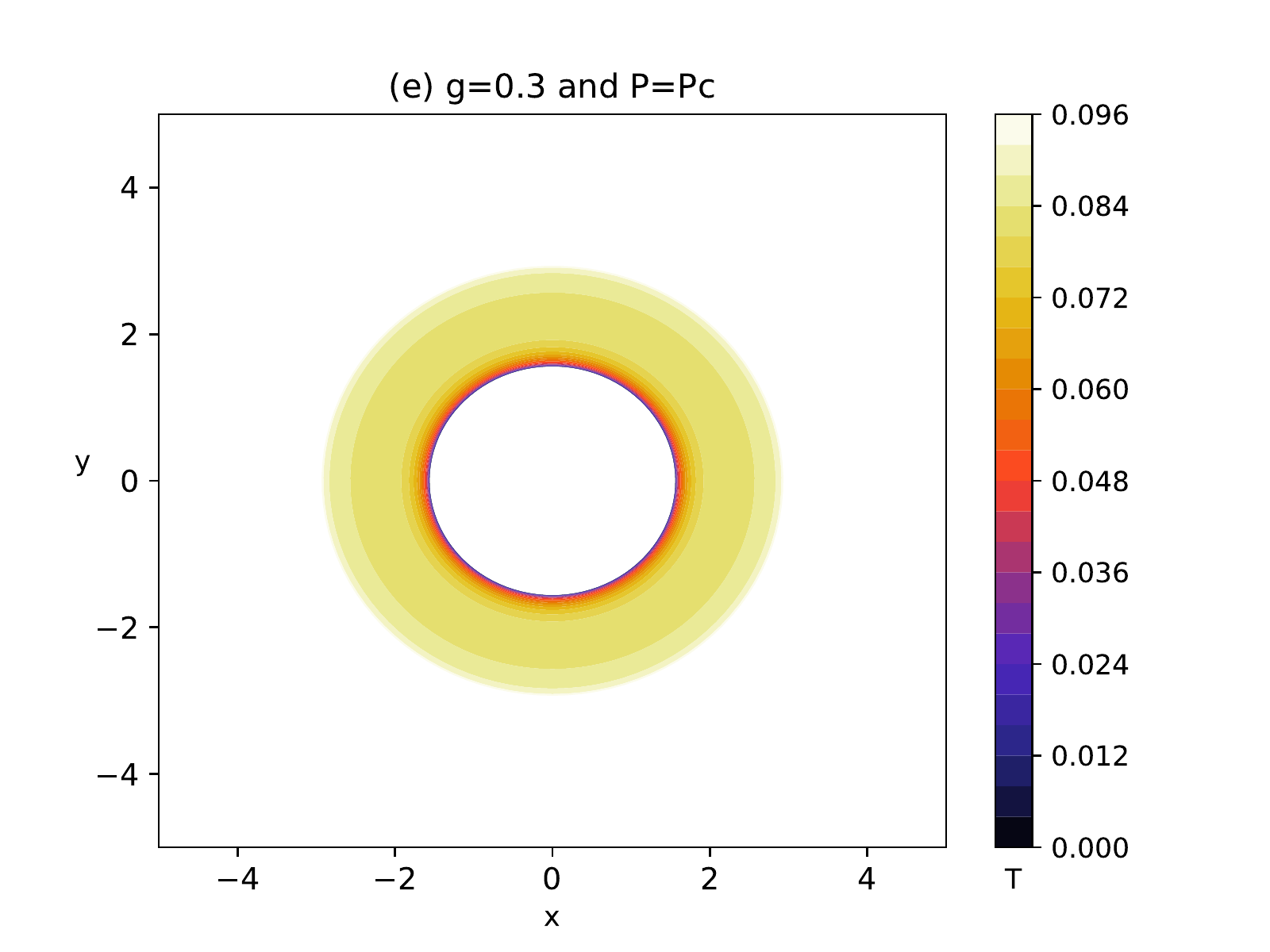}
  \includegraphics[width=5cm,height=4cm]{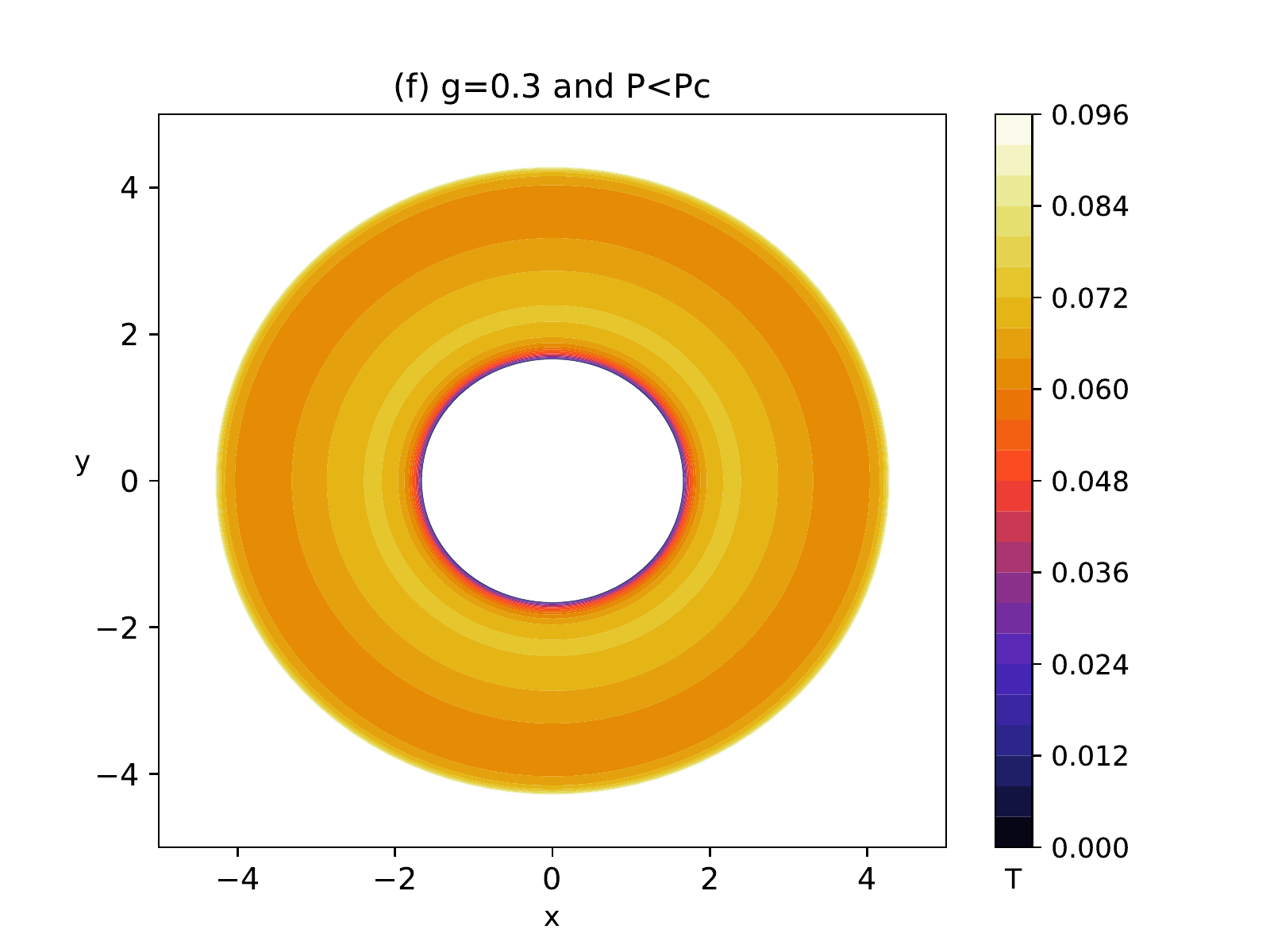}
\parbox[c]{15.0cm}{\footnotesize{\bf Fig~5.}
Thermal profile of the Bardeen-AdS BH for different thermodynamical case. {\em Panel (a)}--$P>P_{\rm c}$ with $g=0.1$, {\em Panel (b)}--$P=P_{\rm c}$ with $g=0.1$ and {\em Panel (c)}--$P<P_{\rm c}$ with $g=0.1$, {\em Panel (d)}--$P>P_{\rm c}$ with $g=0.3$, {\em Panel (e)}--$P=P_{\rm c}$ with $g=0.3$ and {\em Panel (f)}--$P<P_{\rm c}$ with $g=0.3$. The BH mass is taken as $M=60$.}
\label{fig5}
\end{center}

\begin{center}
  \includegraphics[width=6cm,height=5cm]{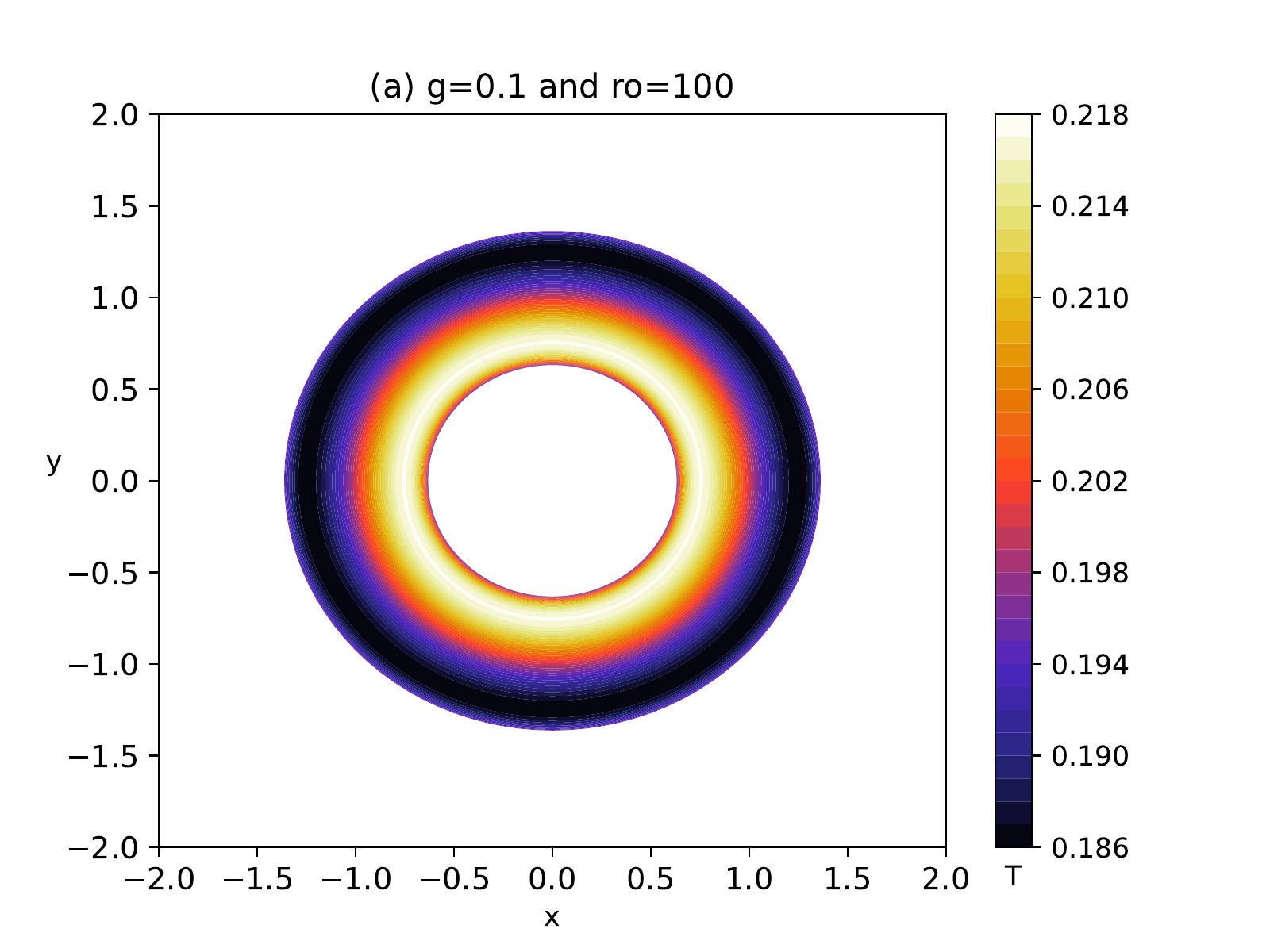}
  \includegraphics[width=6cm,height=5cm]{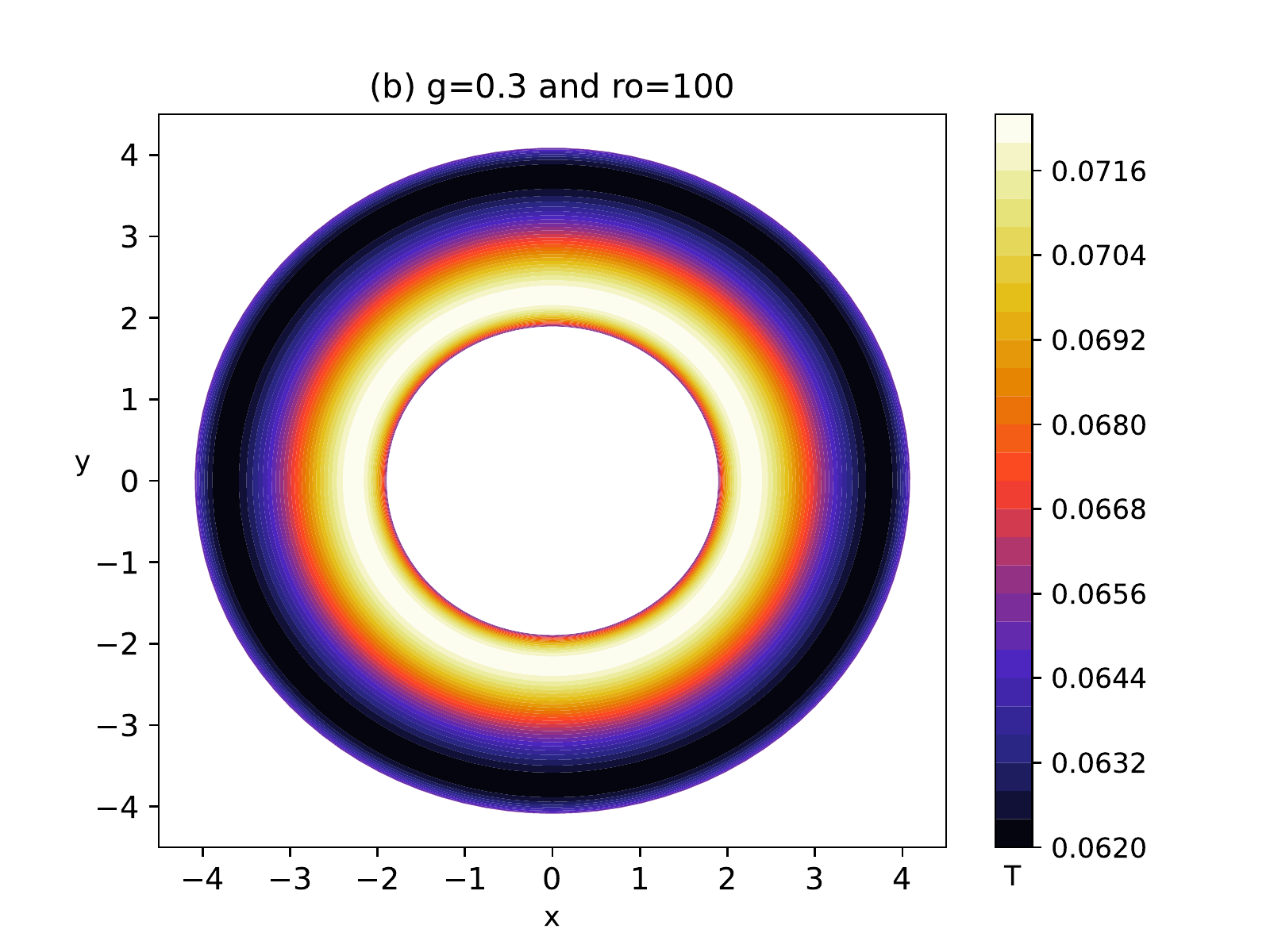}
  \includegraphics[width=6cm,height=5cm]{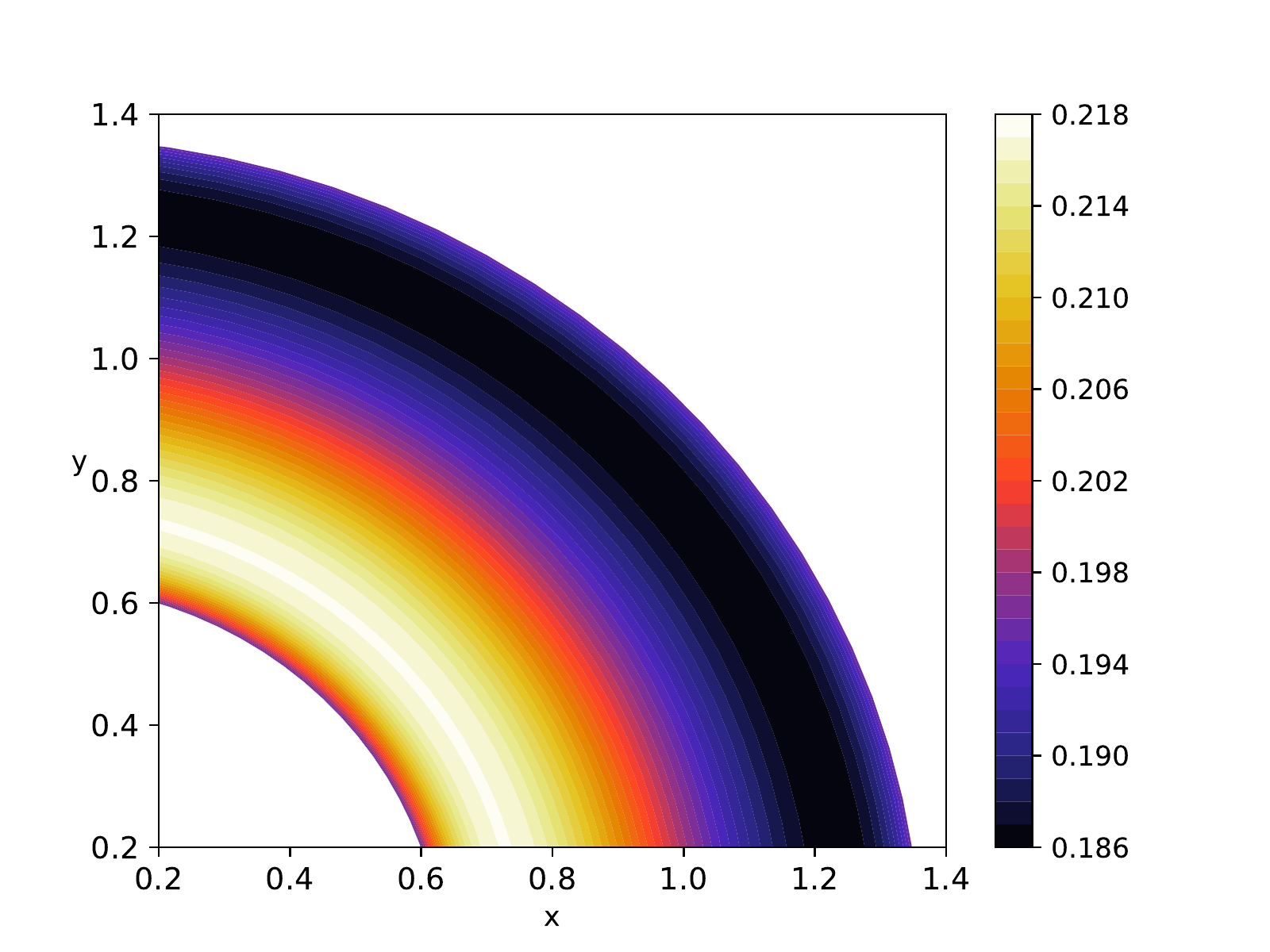}
  \includegraphics[width=6cm,height=5cm]{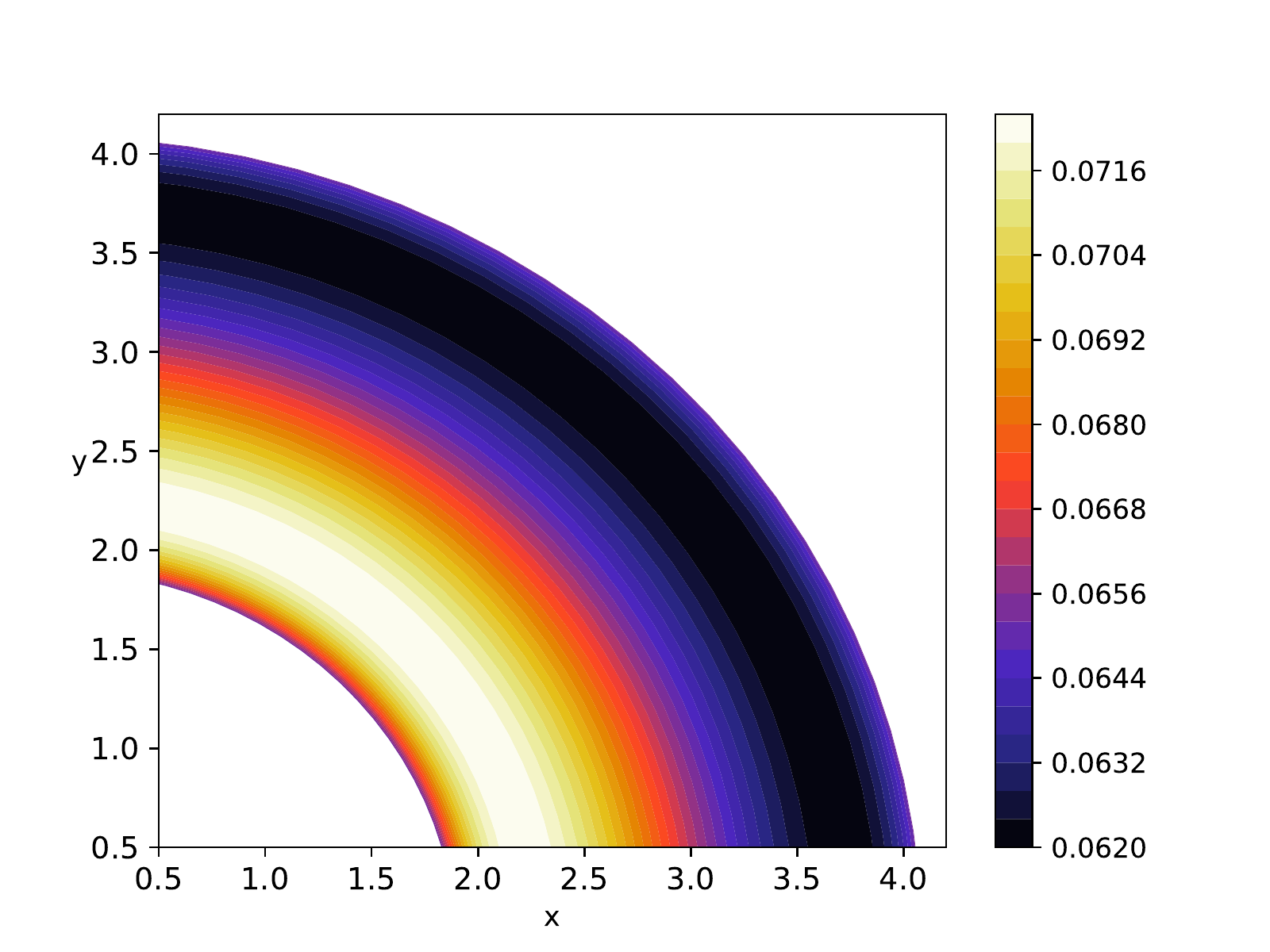}
\parbox[c]{15.0cm}{\footnotesize{\bf Fig~6.}
Thermal profile of the Bardeen-AdS BH for different thermodynamical case. {\em Panel (a)}--$P>P_{\rm c}$ with $g=0.1$, {\em Panel (b)}--$P=P_{\rm c}$ with $g=0.1$ and {\em Panel (c)}--$P<P_{\rm c}$ with $g=0.1$, {\em Panel (d)}--$P>P_{\rm c}$ with $g=0.3$, {\em Panel (e)}--$P=P_{\rm c}$ with $g=0.3$ and {\em Panel (f)}--$P<P_{\rm c}$ with $g=0.3$. The BH mass is taken as $M=60$.}
\label{fig6}
\end{center}

\section{Conclusions and discussions}
\label{sec:5}
\par
The dependence of the BH shadow and thermodynamics in the regular spacetime has been revealed in this analysis. We found that the shadow radius and the event horizon radius display a positive correlation, implying that the BH temperature can be rewritten as a function of the shadow radius in the regular spacetime. For Schwarzschild-AdS BH, the Eq.(\ref{2-20}) is still valid ($g=0$). It can be describe the relationship between the shadow radius and the event horizon radius of the Schwarzschild-AdS BH. We obtained that these two radii still show positive correlation characteristics, while the slope of function curves are greater than regular AdS BH. As the magnetic charge increases, the radius of the event horizon increases, and the corresponding $r_{\rm s}$ get larger. As a result, we believe that the shadow radius can be used to analyze the phase transition structure of the regular AdS BH.

\par
We investigated the phase transition curves in the shadow context. It is found that the shadow radius can replace the event horizon radius to present the Bardeen-AdS BH phase transition process. In the $T-r_{\rm s}$ plane, the BH is the supercritical phase above the critical isobaric, the BH is thermodynamically unstable at the critical pressure. Two-phase transition branches exist below the critical pressure, the $r_{\rm s}$ less than $r_{\rm s1}$ corresponds to a stable small BH, and the $r_{\rm s}$ great than $r_{\rm s2}$ corresponds to a stable large BH. The unstable intermediate branch appears in $(r_{\rm s1},r_{\rm s2})$. By exploring the relationship between the heat capacity and the shadow radius, we found that the shadow radius can reveal the BH phase transition grade. Meanwhile, the equal area law can be constructed with a shadow radius, indicating that the shadow radius may serve as a probe for the phase structure in the regular spacetime.

\par
Utilizing the temperature-shadow radius function, the thermal profile of the Bardeen-AdS BH is established under several representative values of the magnetic charge. We obtained that the size of the BH shadow depends on the pressure. As the pressure decreases, the radius of the thermal profile increases, and the corresponding BH temperature gets smaller. For the $P>P_{\rm c}$, the temperature increases gradually from the center of the shadow to the boundary. The BH is thermodynamically unstable, and the temperature remains constant in the critical region in the $P=P_{\rm c}$ situation. The temperature shows an N-type change trend with the $P<P_{\rm c}$, and its variation law is: increasing $\rightarrow$ decreasing $\rightarrow$ increasing. These results suggest that the regular AdS BH phase transition process can be presented entirely in the thermal profile, and the relationship between the BH shadow and thermodynamics can also be established in the regular spacetime.

\section*{Acknowledgments}
This work is supported by the National Natural Science Foundation of China (Grant No. 11903025).

\clearpage

\end{CJK}
\end{document}